\documentclass[twocolumn]{aastex62}

\usepackage{graphicx}	
\usepackage{amsmath}	
\usepackage{amssymb}	
\usepackage{color}
\usepackage{multirow}

\definecolor{RED}{RGB}{255,0,0}

\accepted{November 10, 2018}

\shorttitle{A case study of TRAPPIST-1\MakeLowercase{g}}
\shortauthors{Wakeford et al.}

\begin{document}

\title{Disentangling the planet from the star in late type M dwarfs: A case study of TRAPPIST-1\MakeLowercase{g}}

\correspondingauthor{Hannah R. Wakeford}
\email{hwakeford@stsci.edu}

\author{H.R. Wakeford}
\affiliation{Space Telescope Science Institute, 3700 San Martin Drive, Baltimore, MD 21218, USA}

\author{N.K. Lewis}
\affiliation{Department of Astronomy and Carl Sagan Institute, Cornell University, 122 Sciences Drive, Ithaca, NY 14853, USA}

\author{J. Fowler}
\affiliation{Space Telescope Science Institute, 3700 San Martin Drive, Baltimore, MD 21218, USA}

\author{G. Bruno}
\affiliation{Space Telescope Science Institute, 3700 San Martin Drive, Baltimore, MD 21218, USA}
\affiliation{INAF‐Osservatorio Astrofisico di Catania, via S. Sofia, 78, 95123 Catania, Italy}

\author{T.J. Wilson}
\affiliation{Space Telescope Science Institute, 3700 San Martin Drive, Baltimore, MD 21218, USA}

\author{S.E. Moran}
\affiliation{Department of Earth and Planetary Sciences, Johns Hopkins University, Baltimore, MD, USA}

\author{J. Valenti}
\affiliation{Space Telescope Science Institute, 3700 San Martin Drive, Baltimore, MD 21218, USA}

\author{N.E. Batalha}
\affiliation{Department of Astronomy and Astrophysics, University of California, Santa Cruz, CA 95064, USA}

\author{J. Filippazzo}
\affiliation{Space Telescope Science Institute, 3700 San Martin Drive, Baltimore, MD 21218, USA}

\author{V. Bourrier}
\affiliation{Observatoire de l'Universit\'{e} de Gen\`{e}ve, 51 chemin des Maillettes, 1290 Sauverny, Switzerland}

\author{S.M. H\"orst}
\affiliation{Department of Earth and Planetary Sciences, Johns Hopkins University, Baltimore, MD, USA}
\affiliation{Space Telescope Science Institute, 3700 San Martin Drive, Baltimore, MD 21218, USA}

\author{S.M. Lederer}
\affiliation{NASA Johnson Space Center, 2101 NASA Parkway, Houston, TX 77058}

\author{J. de Wit}
\affiliation{Department of Earth, Atmospheric and Planetary Sciences, Massachusetts Institute of Technology, Cambridge, MA, USA.}



\begin{abstract}
The atmospheres of late M stars represent a significant challenge in the characterization of any transiting exoplanets due to the presence of strong molecular features in the stellar atmosphere. 
TRAPPIST-1 is an ultra-cool dwarf, host to seven transiting planets, and contains its own molecular signatures which can potentially be imprinted on planetary transit lightcurves due to inhomogeneities in the occulted stellar photosphere. 
We present a case study on TRAPPIST-1g, the largest planet in the system, using a new observation together with previous data, to disentangle the atmospheric transmission of the planet from that of the star. We use the out-of-transit stellar spectra to reconstruct the stellar flux based on one-, two-, and three-temperature components. We find that TRAPPIST-1 is a 0.08 M$_*$, 0.117 R$_*$, M8V star with a photospheric effective temperature of 2400\,K, with $\sim$35\% 3000\,K spot coverage and a very small fraction, $<$3\%, of $\sim$5800\,K hot spot. We calculate a planetary radius for TRAPPIST-1g to be R$_p$\,=\,1.124\,R$_\oplus$ with a planetary density of $\rho_p$\,=\,0.8214\,$\rho_\oplus$.
Based on the stellar reconstruction there are eleven plausible scenarios for the combined stellar photosphere and planet transit geometry; in our analysis we are able to rule out 8 of the 11 scenarios. Using planetary models we evaluate the remaining scenarios with respect to the transmission spectrum of TRAPPIST-1g. We conclude that the planetary transmission spectrum is likely not contaminated by any stellar spectral features, and are able to rule out a clear solar H$_2$/He-dominated atmosphere at greater than 3-sigma. 
\end{abstract}



\section{Introduction} \label{sec:intro}
Planets around M dwarf stars represent a rich opportunity for the study of exoplanet atmospheres that comes with a few challenges. Because M dwarfs are smaller than their solar type cousins, signals from the planets that orbit closely to them are significantly stronger. However, the atmospheres of M dwarf stars are also significantly more complex than solar type stars. Stellar variability, due to spots and other stellar atmospheric features, is a common feature of M dwarf star atmospheres, which inject potential false positive signals into observations of the planets in these systems. Late type M dwarfs (M5-M8) present particular challenges as they can contain molecules like water vapor and form aerosols in their atmospheres \citep{Allard2012}, which are key features of cooler planetary atmospheres. In order to take advantage of the rich opportunities offered by the study of planets in orbit around M dwarfs, we must develop robust techniques to disentangle planetary and stellar signals.

The TRAPPIST-1 system has been the subject of many investigations since the discovery of seven Earth-sized worlds orbiting this ultra-cool M dwarf \citep{Gillon2016,gillon2017}. TRAPPIST-1 is an M8 star, with a radius of R$_*$\,=\,0.121\,R$_\odot$, a mass of M$_*$\,=\,0.089\,M$_\odot$, and T$_{eff}$\,=\,2511\,K \citep{vangrootle2018}. A global analysis of all \textit{Spitzer} photometry of this system was presented in \citet{delrez2018} which further refined the parameters for each of the planets from the original studies. Specifically, this \textit{Spitzer} analysis was performed using the planetary radii calculated using the updated stellar properties from \citet{vangrootle2018}. This was followed by long-term TTV analysis using both \textit{Spitzer} and \textit{K2} photometry to provide precise constraints on the masses, and therefore the densities, of each planet in the system \citep{Grimm2018}. 
 
Follow-up observations characterizing the atmospheres of TRAPPIST-1b-g have previously been presented in \citet{deWit2016,dewit2018}. These results show that the atmospheres of TRAPPIST-1b-f do not show evidence for H/He dominated atmospheres that are free from clouds. However, a H/He-dominated atmosphere for TRAPPIST-1g could not be ruled out due to large uncertainties on the transmission spectrum. Further analysis in \citet{Moran2018} demonstrated that photochemical hazes and large cloud opacities in hydrogen-rich atmospheres are also unlikely for TRAPPIST-1~d and e, however updated mass constraints \citep{Grimm2018} prohibit the confident exclusion of a cloud-free hydrogen-rich atmosphere for TRAPPIST-1~f. Constraints on TRAPPIST-1g were not possible given the data presented in \citet{dewit2018}. 

Here we use the most up-to-date system parameters for TRAPPIST-1 from the literature, to investigate the atmosphere of TRAPPIST-1g the 6th planet out from the star and the largest of the TRAPPIST-1 planets. In \S\,\ref{sec:obs} and \ref{sec:analysis} we detail the new observations, limb-darkening calculations, and lightcurve analysis conducted to produce a measured transit depth with wavelength. In \S\ref{sec:stellar} and \ref{sec:transmission_fit} we introduce the star TRAPPIST-1 with respect to stellar activity and discuss the impact the stellar flux will have on the measured transmission spectrum. In these sections we go into detail about the stellar \textit{contrast effect} (e.g., \citealt{Cauley2018}), and the contamination fraction associated with portions of the star being different temperatures (e.g., \citealt{rackham2018}). We use the out-of-transit spectra to fit for these effects and apply it to our measured transmission spectrum to explore the potential planetary absorption signatures. In \S\ref{sec:interp} we interpret the planetary transmission spectrum using atmospheric models, and in \S\ref{sec:end} we summarize our results obtained for the star and planet.  

\begin{table}
\centering
\caption{Compiled measurements of the TRAPPIST-1g system parameters from the literature. In this analysis we update parameters for the star TRAPPIST-1 in Table \ref{table:star_params}, and the planet radius and density in \S\ref{sec:end}.}
\begin{tabular}{p{0.26\textwidth}p{0.08\textwidth}p{0.09\textwidth}}
\hline
\hline
Parameter & Value & Uncertainty\\
\hline
\multicolumn{3}{c}{\textbf{Star}}\\
Radius, R$_*$ (R$_\odot$)$^a$ & 0.121 & $\pm$0.003\\
Mass, M$_*$ (M$_\odot$)$^a$  & 0.089 & $\pm$0.006\\
Density, $\rho_*$ ($\rho_{\odot}$)$^b$ & 51.1 & $^{+1.2}_{-2.4}$\\
Effective Temperature, T$_{eff}$ (K)$^b$ & 2511 & $\pm$37\\
Luminosity, L$_*$ (L$_\odot$)$^a$ & 5.22$\times$10$^{-4}$ & $\pm$0.19$\times$10$^{-4}$\\
Metallicity, $[$Fe/H$]$ (dex)$^b$ & 0.04 & $\pm$0.08\\
Gravity, log(g)$^c$  & 5.22 &$\pm$0.08\\
Age (Gyr)$^d$ & 7.6 & $\pm$2.2 \\
Parallax, $\pi$ (mas)$^e$ & 80.451 & $\pm$0.121 \\
Distance, d (pc)$^e$ & 12.430 & $\pm$0.019 \\
\hline
\multicolumn{3}{c}{\textbf{Planet}}\\
Radius, R$_p$ (R$_\oplus$)$^b$ & 1.154 & $\pm$0.029\\
Mass, M$_p$ (M$\oplus$)$^f$ & 1.148 & $^{+0.098}_{-0.095}$\\
Density, $\rho_p$ ($\rho_\oplus$)$^f$ & 0.759 & $^{+0.033}_{-0.034}$\\
Inclination, i ($^\circ$)$^b$ & 89.721 & $^{+0.019}_{-0.026}$\\
Eccentricity, e$^f$ & 0.00208 & $\pm$0.00058\\
Argument of periapsis, $\omega$ ($^\circ$)$^d$ & 191.34 & $\pm$13.83 \\
Period, P (days)$^b$ & 12.35447 & $\pm$0.000018\\
Semi-major axis, a (AU)$^f$ & 0.04687692 & $\pm$3.2$\times$10$^{-7}$\\
a/R$_*$ $^b$ & 83.5 & $^{+0.7}_{-1.3}$\\
Gravity, g$_p$ (ms$^{-2}$)$^f$ & 7.4302 & $^{+0.71}_{-0.65}$\\
Equilibrium Temperature, T$_{eq}$ (K)$^b$ & 194.5 & $\pm$2.7\\
\hline
\multicolumn{3}{l}{a - \citet{vangrootle2018};~ b - \citet{delrez2018}}\\
\multicolumn{3}{l}{c - \citet{Filippazzo2015};~ d - \citet{Burgasser2017}}\\
\multicolumn{3}{l}{e- \citet{gaia2018};~ f - \citet{Grimm2018}}\\
\end{tabular}
\label{table:system_params}
\end{table}

\section{TRAPPIST-1\MakeLowercase{g} Observations} \label{sec:obs}
We observed one transit of TRAPPIST-1g with the Hubble Space Telescope (HST) Wide Field Camera 3 (WFC3) G141 grism on 10 December 2017 as part of GO-15304 (PI de Wit). We obtained 60 exposures in spatial scan mode over the course of four HST orbits each with an exposure time of 112 seconds. The scan rate was set to 0.02 arcseconds/second, resulting in a scan covering $\sim$17 pixels in the cross-dispersion direction. We use the \emph{ima} format from the \texttt{calwf3} pipeline to extract the stellar spectra and analyse the transit lightcurves. In addition to this new observation, we re-analyse the previously published TRAPPIST-1g observations \citep{dewit2018} from GO-14873 visit 2, which captured the ingress of the planet. We choose to ignore the visit 4 observations from the same program as they experienced pointing issues following the crossing of HST into the South Atlantic Anomaly (SAA). As the GO-14873 visit 4 measurements were not of high enough quality to add definitive measurements to the final results we choose to leave them out of this combined analysis. Hereafter, the original GO-14873 observation will be called transit 1, and the new GO-15304 observation will be called transit 2.

For each exposure, in each transit, we flag and remove any cosmic rays incident on the detector by stacking each exposure in time and replacing any pixel value greater than 5-$\sigma$ from the median of the array to the median value \citep{Nikolov2014}. We also check each flagged pixel for spatial variation within the exposure by comparing the flux to the horizontal pixel flux values within 5 pixels of the flagged pixel. This spatial check is used to avoid correction applied to small variations within the spectral trace due to inconsistent scan rates. We perform this check iteratively until no more cosmic rays are flagged; in each of these observations this required two iterations of our cosmic ray flagging. 

We extract the stellar spectrum from each exposure using an optimized aperture determined by taking the out-of-transit exposures and minimizing the standard deviation (see \citealt{wakeford2016}). For transit 1 we use an aperture of 31 pixels in the cross dispersion direction, and for transit 2 we use an aperture of 29 pixels. We also extract the stellar spectrum using the differencing method outlined in \citet{Evans2016}, where the stellar spectrum is reconstructed from the individual non-destructive reads in each exposure and setting all external values which do not contain the stellar spectrum to zero via a top-hat filter centered on the exposed portion of the detector. We find that both of these methods produce the same standard deviation in the out-of-transit spectra and therefore either method is appropriate for further analysis of the lightcurves. 

\section{Measuring the Transit depth}\label{sec:analysis}
For each observation we analyse the broadband lightcurve by summing up the stellar flux in each exposure between 1.1--1.7\,$\mu$m. We first analyse the two transits separately using two different methods: one by marginalizing over a series of systematic models using a least-squares minimizer \citep{wakeford2016}, and two using an Markov chain Monte Carlo (MCMC) with a single exponential ramp model to account for the instrumental systematics (\citealt{bruno2018} and references therein). From these we achieve similar transit depths well within the uncertainties for each transit. In each analysis method we fix the values for inclination, a/R$_*$, period, and R$_*$ to the most up-to-date values in the literature (see Table\,\ref{table:system_params}), and fit for the R$_p$, baseline stellar flux, and center of transit time. We attempted to fit for all the system parameters with our data but found non-significant differences from the literature values, as well as the transit depths, therefore we keep them fixed in this analysis. We find a center of transit time of 2457751.81200$\pm$5$\times$10$^{-5}$\,BJD$_\mathrm{TBD}$ for transit 1 and 2458097.72346$\pm$4$\times$10$^{-4}$\,BJD$_\mathrm{TBD}$ for transit 2, and a broadband transit depth for TRAPPIST-1g of 7719$\pm$86\,ppm and 7697$\pm$52\,ppm respectively. Transit times were converted from MJD to BJD$_\mathrm{TBD}$ following \citet{Eastman2010}. 
We fix limb-darkening values to those computed for a stellar model of T$_\mathrm{eff}$\,=\,2663\,K, log(g)\,=\,5.22, and [Fe/H]\,=\,0.04 (see \S\ref{sec:ld} for details). We do not observe any evidence of stellar spot crossings in either transit lightcurve, or in our spectroscopic fits (see discussion on TRAPPIST-1 the star \S\ref{sec:stellar}).

\begin{figure*}
\centering
	\includegraphics[width=0.95\textwidth]{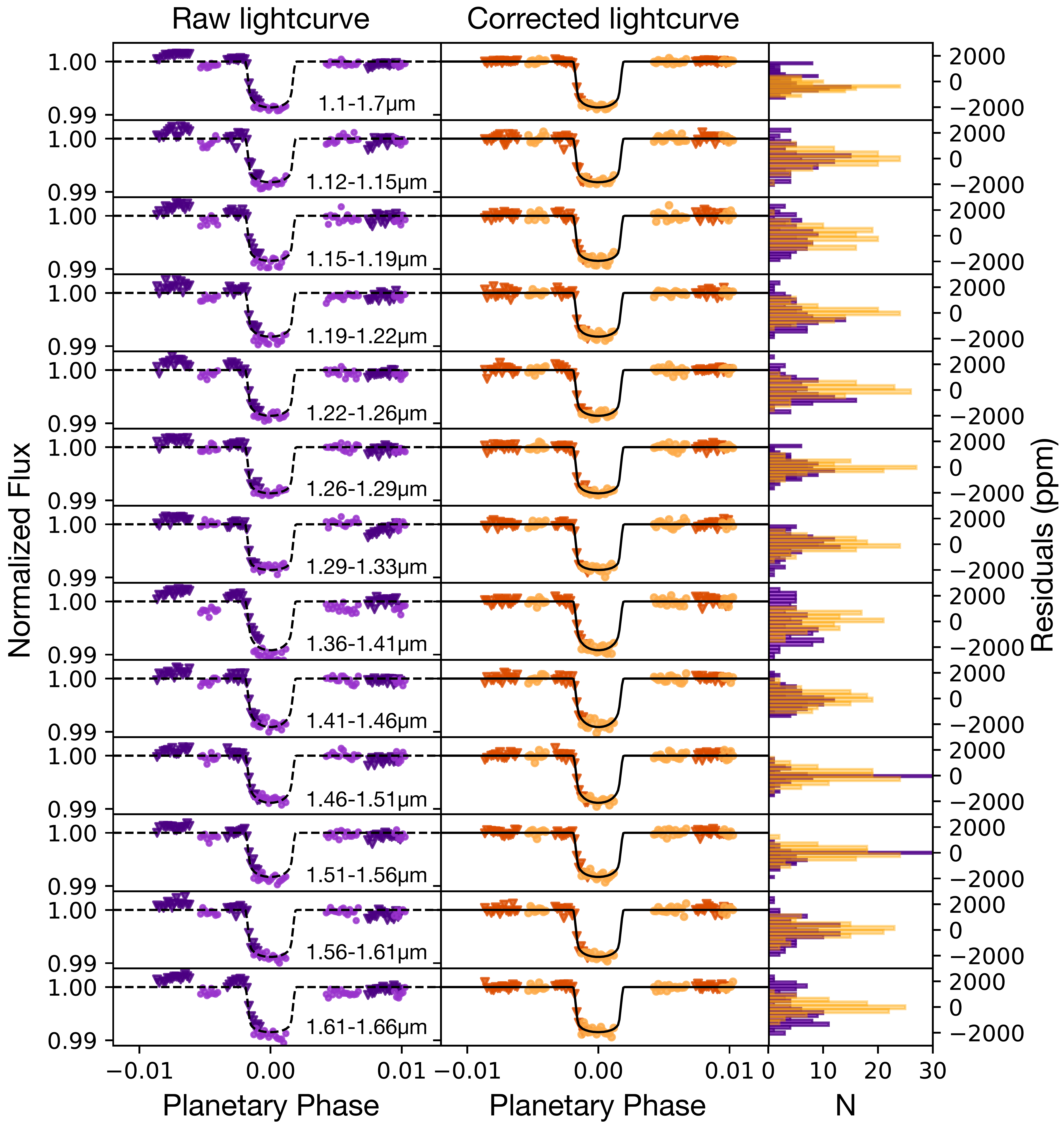}
    \caption{Left: Raw spectroscopic lightcurves for transit 1 (dark) and 2 (light) in terms of planetary phase.  Middle: Corrected spectroscopic lightcurves for transit 1 (dark) and 2 (light). Right: Histogram of the raw residuals (dark purple) and corrected residuals (light orange). The broadband lightcurve is shown in the top panel of the figure.}
    \label{fig:T1g_t1}
\end{figure*}

In addition to fitting each transit independently, we conduct a joint fit of the two transits together fitting for a common planetary radius across both observations. This follows the assumption that the star is not changing between observations (see section \ref{sec:star_fit}). We show the broadband lightcurve for both transits fit together in the top panel of Fig.\,\ref{fig:T1g_t1}. In our joint fit we find a transit depth of 7736$\pm$35\,ppm and the same transit times as in the individual fits but with marginal reductions in the uncertainties. In the joint fits when marginalizing over a grid of potential systematic models to correct the lightcurves we find a best fit systematic model, $S$ \citep{wakeford2016}, with the form,
\begin{equation}
S = T_1\theta \times \sum^3_{i=1}p_i\phi^i
\end{equation}  
where $\theta$ is the planetary phase, $\phi$ is the HST orbital phase with the length of a HST orbit set to 95.25 minutes, and $T_1$ and $p_i$ are free parameters. We discuss the wavelength dependence of these in \S\ref{sec:wavelength}. This model is applied separately to each transit array and functionally corrects for a linear slope in time across the whole transit and a third order polynomial in HST phase to account for ``HST breathing'' effect{}s due to thermal variations of the telescope throughout the HST orbit. 

\subsection{Limb-darkening}\label{sec:ld}
For each lightcurve we analyse, we determined limb-darkening coefficients using the online ExoCTK\footnote{exoctk.stsci.edu/limb\_darkening} limb-darkening tool. This tool fits a variety of limb darkening functions to theoretical specific intensity spectra ($I$) of the Phoenix ACES model atmosphere grid downloaded from the G$\ddot{o}$ttingen spectral library \citep{Husser2013}. 

To determine the limb darkening coefficients for the star, the tool interpolates the model grid to the stellar parameters of TRAPPIST-1 (Table \ref{table:system_params}). A specific intensity spectrum is generated for each value of $\mu$ using the \texttt{linear} method of the \texttt{scipy.interpolate} package \citep{scipy} \texttt{RegularGridInterpolator} class. This is then resampled to and multiplied by the throughput of the WFC3 IR G141 grism.

\begin{figure}
\centering
	\includegraphics[width=0.45\textwidth]{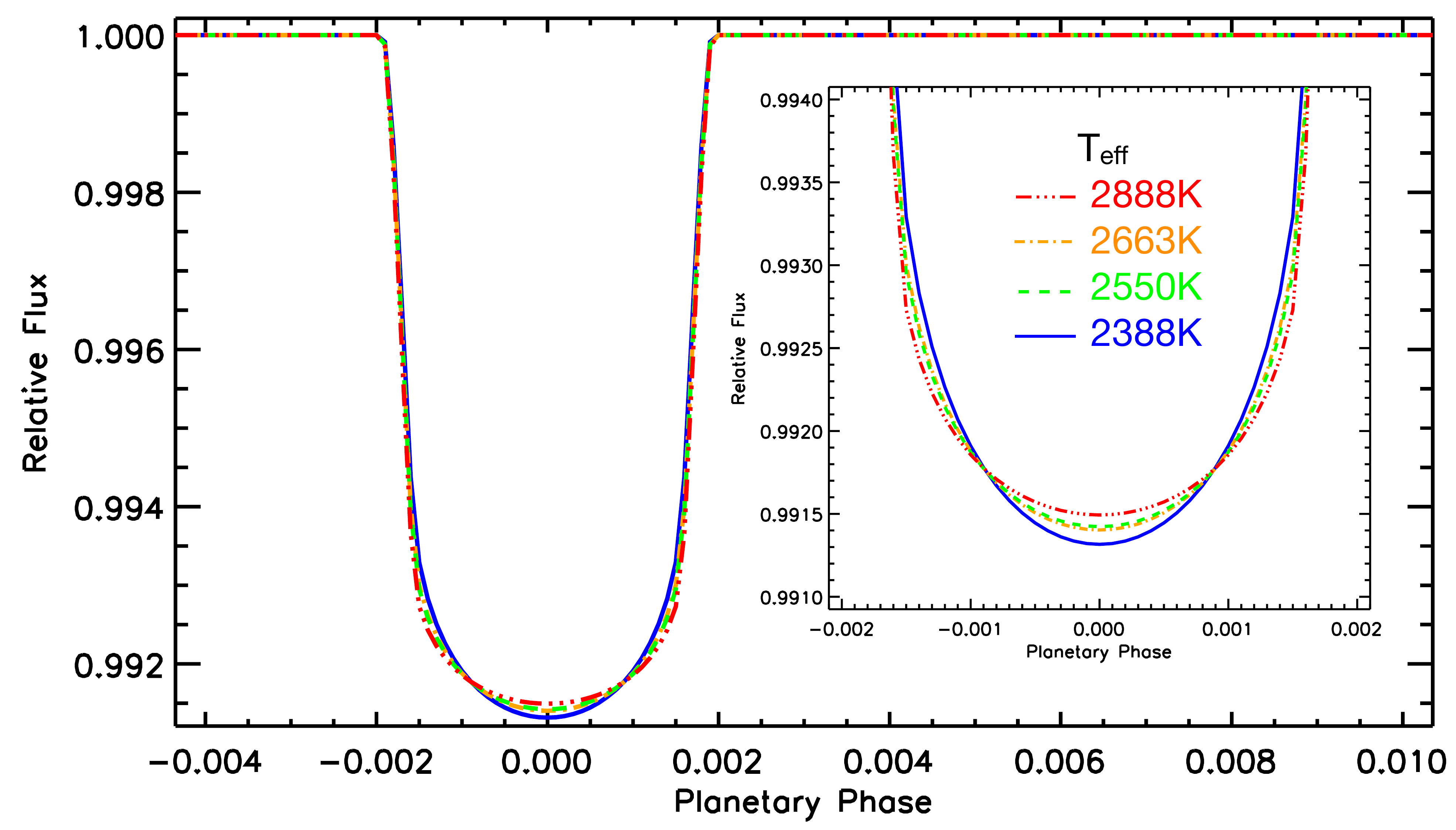}
    \caption{Differences in the transit profile for TRAPPIST-1g due limb darkening for different T$_{eff}$s. As the effective stellar temperature increases the transit profile becomes shallower and wider.} \label{fig:LD_check}
\end{figure}

When fitting limb-darkening functions, PHOENIX $\mu$ values should be scaled to yield $\mu$'\,=\,0 at the stellar radius. We define $\mu$'\,=\,($\mu$ - $\mu_0$)/(1 - $\mu_0$), where I/I$_c$ = 0.01 at $\mu$ = $\mu_0$. The value of $\mu_0$ is a function of wavelength. The mean specific intensity at each $\mu$ is calculated ($\overline{I}_{\mu}$) and then divided by the mean specific intensity at $\mu=1$ to determine the relative limb darkening at each angle. This produces a curve that goes from 0 at the stellar limb to 1 at the disk center. The Levenberg-Marquardt method of the \texttt{scipy.optimize.curve\_fit} function is then used to find the best fit coefficients for the desired limb darkening function. For this analysis, we fit two commonly used functional forms for limb darkening, the quadratic rule and the 4-parameter law (see \citealt{deWit2016}).
The presented transit depths are computed using the 4-parameter limb-darkening law, however, we list both the quadratic and 4-parameter coefficients for each wavelength channel in Table \ref{table:transmission_spec}. The 4-parameter limb-darkening law is chosen in this analysis as this has been demonstrated to avoid biases in the data better than the quadratic law \citep{Espinoza2015}. 
\begin{table*}[ht]
\centering
\caption{Transit depths computed for TRAPPIST-1g for transit 1 and 2 separately as well as a joint fit across both transits. Limb-darkening parameters for T$_\mathrm{eff}$ = 2663\,K are shown for each wavelength range.}
\begin{tabular}{cc|cc|cc|cc|cccccc}
\hline
\hline
& & \multicolumn{2}{ c| }{Transit 1} & \multicolumn{2}{ c| }{Transit 2} & \multicolumn{2}{ c| }{Joint Fit} & \multicolumn{6}{ c }{Limb-darkening coefficients} \\
Central $\lambda$ & $\Delta\lambda$ & $D_M$ & $\sigma D_M$ & $D_M$ & $\sigma D_M$ & $D_M$ & $\sigma D_M$ &  c1 & c2 & c3 & c4 & u1 & u2 \\
 ($\mu$m) & ($\mu$m) & (ppm) & (ppm) & (ppm) & (ppm) & (ppm) & (ppm) & \multicolumn{4}{ c| }{4-parameter} & \multicolumn{2}{ c }{Quadratic}\\
\hline
1.4000 & 0.3000 & 7719 & 86 & 7697 & 52 & 7736 & 35 & 1.377 & -0.799 & 0.332 & -0.064 & 0.130 & 0.427\\
\hline
1.1375 & 0.0175 & 7363 & 338 & 7370 & 237 & 7520 & 150 & 1.570 & -0.977 & 0.382 & -0.060 & 0.113 & 0.473 \\
1.1725 & 0.0175 & 7479 & 321 & 7645 & 206 & 7630 & 136 & 1.718 & -1.529 & 0.909 & -0.234 & 0.089 & 0.430 \\
1.2075 & 0.0175 & 7749 & 326 & 7388 & 184 & 7669 & 137 & 1.568 & -1.295 & 0.739 & -0.186 & 0.096 & 0.413 \\
1.2425 & 0.0175 & 7643 & 281 & 7698 & 209 & 7776 & 124 & 1.504 & -1.194 & 0.666 & -0.165 & 0.099 & 0.405 \\
1.2775 & 0.0175 & 8198 & 306 & 8018 & 200 & 7872 & 125 & 1.504 & -1.340 & 0.810 & -0.213 & 0.083 & 0.378 \\
1.3125 & 0.0175 & 8093 & 262 & 7878 & 167 & 7765 & 114 & 1.723 & -1.598 & 0.959 & -0.248 & 0.073 & 0.419 \\
1.3850 & 0.0250 & 7587 & 299 & 7920 & 173 & 8011 & 114 & 1.089 & 0.047 & -0.409 & 0.169 & 0.179 & 0.468 \\
1.4350 & 0.0250 & 8203 & 281 & 7665 & 190 & 7730 & 116 & 0.817 & 0.395 & -0.491 & 0.155 & 0.265 & 0.418 \\
1.4850 & 0.0250 & 8128 & 281 & 7457 & 174 & 7778 & 121 & 1.007 & 0.053 & -0.318 & 0.123 & 0.203 & 0.434 \\
1.5350 & 0.0250 & 6918 & 291 & 7423 & 154 & 7335 & 111 & 1.173 & -0.168 & -0.244 & 0.122 & 0.159 & 0.461 \\
1.5850 & 0.0250 & 7884 & 268 & 7924 & 136 & 7977 & 109 & 1.440 & -0.982 & 0.472 & -0.104 & 0.108 & 0.419 \\
1.6350 & 0.0250 & 7984 & 244 & 7523 & 143 & 7754 & 110 & 1.492 & -1.219 & 0.682 & -0.170 & 0.086 & 0.396 \\
\hline
\end{tabular}
\label{table:transmission_spec}
\end{table*}
Limb-darkening is highly dependent on the stellar models used to compute the coefficients. To demonstrate this effect we show the transit profile for a nominal transit of TRAPPIST-1g in Fig.\,\ref{fig:LD_check} for a series of different stellar profiles. This clearly demonstrates that as the temperature of the star increases, the shape of the transit profile becomes shorter and wider. Given nominal phase coverage of the transit profile the data can be used to place constraints on the limb-darkening coefficients. To test this we allow the limb-darkening parameters to be freely fit in our broadband lightcurve analysis and find that the fixed parameters are well within the resultant posteriors \citep{bruno2018} for limb-darkening coefficents computed for T$_\mathrm{eff}$\,=\,2666\,K. We additionally computed the transmission spectrum for TRAPPIST-1g (described in the following section) for each profile in Fig.\,\ref{fig:LD_check} and find that it has negligible impact on the absolute depth of the measured transit and no effect on the shape of the measured transmission spectrum. This test suggests that the stellar profile chosen for these parameters was a reasonable approximation to a uniform stellar photosphere (see \S\ref{sec:star_fit}).

\subsection{Spectroscopic Analysis}
\begin{figure}
\centering
	\includegraphics[width=0.48\textwidth]{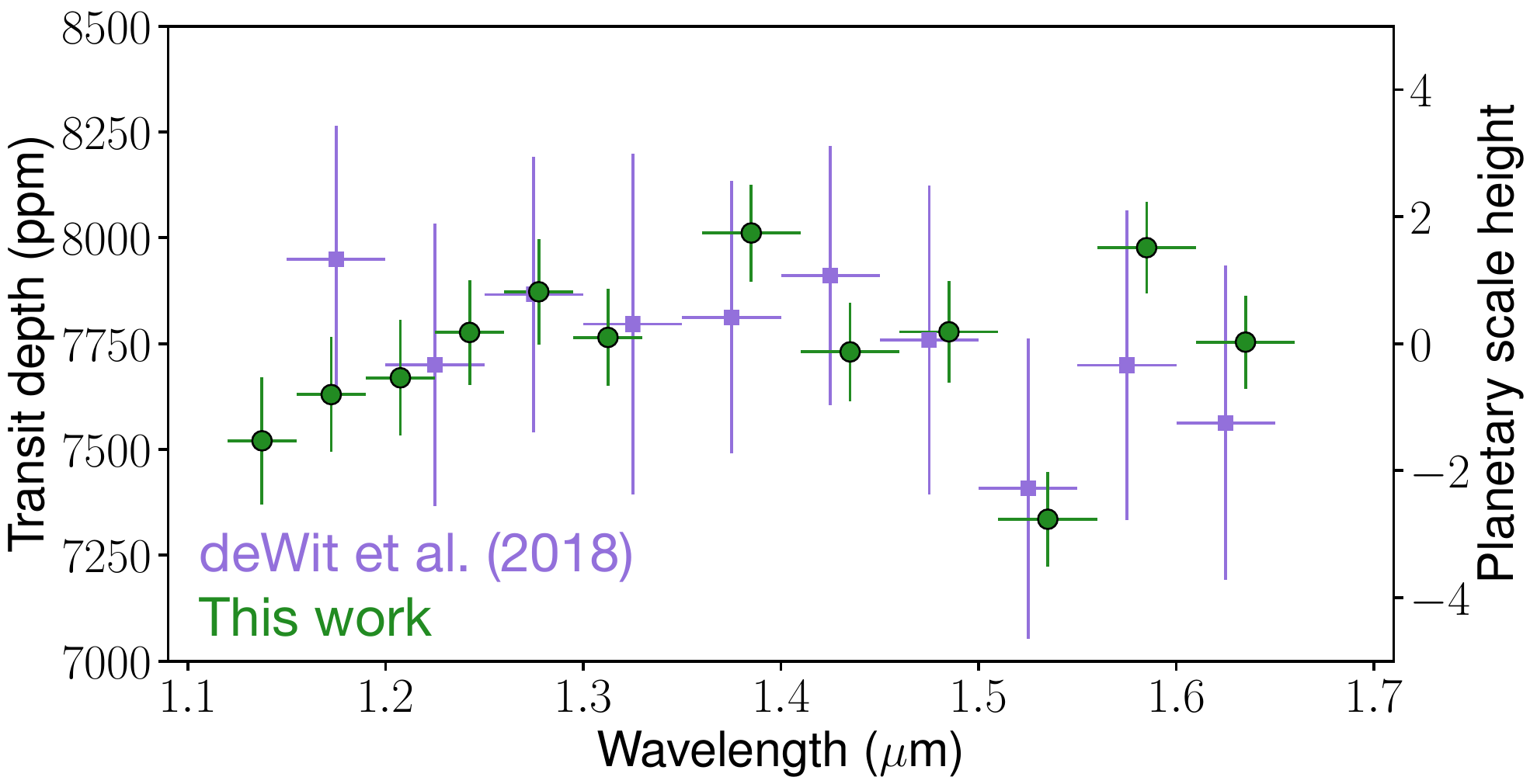}
    \caption{Measured transmission spectrum (green circles) from a joint fit of the two transit events (see Table \ref{table:transmission_spec}), against the previously published transmission spectrum from \citet{dewit2018} (purple squares).}
    \label{fig:trans_spec}
\end{figure}

To measure the wavelength dependent transmission spectrum we divide the stellar spectrum into a series of spectroscopic channels. In this analysis we fix the center of transit times to those obtained from the broadband analysis and keep all previous values for a/R$_*$, inclination, and period fixed. We again analyse each spectroscopic lightcurve using two different methods, using marginalization across a series of systematic models and an MCMC analysis,  for each transit independently, and using a joint fit across both transits. We fix the limb-darkening values to those computed for a stellar model of T$_\mathrm{eff}$\,=\,2663\,K (see \S\ref{sec:star_fit} and \ref{sec:transmission_fit} for more details), log(g)\,=\,5.2, and [Fe/H]\,=\,0.04 in each wavelength bin (see \S\ref{sec:ld} for details on the limb-darkening). The results of each lightcurve fit and the limb-darkening coefficients for both quadratic and four-parameter limb-darkening laws are listed in Table\,\ref{table:transmission_spec}. We show each spectroscopic lightcurve in their uncorrected and corrected form in Fig. \ref{fig:T1g_t1}, along with the residuals across both transits. 
The final measured transmission spectrum from a joint fit across both transits is shown in Fig.\,\ref{fig:trans_spec} with the original published transmission spectrum in \citet{dewit2018} which just looked at transit 1. In this analysis we significantly reduce the uncertainties by combining the two transit events, which when combined give better phase coverage of the transit. The measured transmission spectrum, while showing some structure, is well within the original measurements, further validating the initial measurements presented. 

\subsubsection{Wavelength Dependences}\label{sec:wavelength}
The analysis of the spectroscopic lightcurves required additional attention as the lightcurves exhibit wavelength dependent systematics which mean that a broadband correction using the residuals of the white lightcurve fit cannot be used to fully correct each spectroscopic lightcurve (e.g., \citealt{Deming2013,Stevenson2014,bruno2018}). Instead, we fit for the systematic model in each wavelength bin (e.g., \citealt{wakeford2016}) to correct each lightcurve for instrumental or observation imposed systematics. 

The wavelength dependence of the systematics in these cases is likely due to the change in the total flux received by each pixel across the stellar spectrum, due to the presence of broad absorption features in the star. These stellar absorption features vary the flux of the star across the wavelength direction from a maximum of 26,000 electrons/pixel down to 16,000 electrons/pixel in transit 1, and from 34,000--22,000 electrons/pixel in transit 2. This change in the flux on the detector introduces a varied systematic ramp across each spectroscopic bin, which has been previously noted in \citet{Wilkins2014} and \citet{Wakeford2017}, associated with the per pixel count levels. This changing effect can be easily accounted for by allowing the systematic model to be redefined by each lightcurve as is done with the broadband lightcurve. 

Additionally, it is important to note the wavelength dependence of limb-darkening coefficients for TRAPPIST-1, which changes the shape of the lightcurve across the stellar absorption features. This can most clearly be seen in Fig. \ref{fig:T1g_t1} in the shape of the lightcurve for the 1.29--1.33\,$\mu$m bin which is outside of the water band of the star and the 1.36--1.41\,$\mu$m bin which is inside the water absorption band of the star. Outside of the band the effects of limb-darkening are greatly reduced compared to inside the band. This specifically affects the curvature of the lightcurve and the absolute depth. It is therefore important to use limb-darkening coefficients specifically calculated for each wavelength bin to account for changes in the star.  We list the limb-darkening coefficients for each of our spectroscopic bins in Table \ref{table:transmission_spec} using the 4-parameter and quadratic limb-darkening laws.\\

In summary, we have extracted and analysed two transits of TRAPPIST-1g for WFC3 G141 grism spectroscopy. We use both a marginalization method \citep{wakeford2016} and MCMC analysis \citep{bruno2018} to measure the transit depth as a function of wavelength on both transits separately, and in a joint analysis, to determine the measured transmission spectrum. In each spectroscopic lightcurve analysis we allow the systematic model to be fit independently as there are apparent wavelength dependent systematic effects in the lightcurves due to changing flux across the stellar spectrum. The resultant measured transmission spectrum combining both transit observations lies well within the uncertainties of the previously published results in \citet{dewit2018}, but has reduced uncertainties, from $\sim$340\,ppm to $\sim$120\,ppm, due to the added information from transit 2 and the application of a joint fit. In the following sections we discuss how the measured transmission spectrum and the planetary transit can be used to better approximate the composition of the stellar photosphere and disentangle the spectrum of the star from the planet. 

\section{TRAPPIST-1, The Star} \label{sec:stellar}
Before we can interpret the transmission spectrum we need to further explore the nature of TRAPPIST-1 over our observed wavelength range (1.1--1.7\,$\mu$m). The effect of stellar activity on transmission spectra is not a new phenomenon \citep{sing2011b,huitson2013,Cauley2017,bruno2018,Cauley2018}, where inhomogeneities in the stellar flux can cause noticeable effects on the measured transit depths. In addition, the time variable nature of stellar activity can have an impact on transmission spectra produced from observations obtained from different epochs (e.g., \citealt{Zellem2017}). Here we adopt the term \textit{contrast effect} from \citet{Cauley2018} to describe the effect that differences in the spectral signatures of various stellar regions -- active or otherwise -- has on the bulk point source properties of the star.

Recent studies have more specifically explored the impact of stellar contamination from TRAPPIST-1 on the measured spectra of its transiting planets (e.g., \citealt{rackham2018,zhang2018,Ducrot2018}). The studies conducted in \citet{rackham2018} and \citet{zhang2018} find that molecular features in the combined transmission spectra for six of the planets in the TRAPPIST-1 system, detailed in \citet{deWit2016,dewit2018}, can predominantly be explained by inverse absorption features in the star.
Specifically, \citet{rackham2018} explored the impact that an inhomogeneous stellar surface flux would have on transmission spectral studies for M stars, detailing that inhomogeneities on the stellar surface will directly translate into the features on the transmission spectrum which can be then confused with planetary features. In their paper they outline a simple analytical formula to fit for the contrast effect on the stellar flux; this has since been incorporated into more complex retrieval techniques on multi-wavelength observations of giant planets \citep{Pinhas2018}. However, even with high precision planetary transit observations the constraints on the stellar atmosphere are minimal. Therefore, for this investigation on a smaller planet with a smaller cooler star, this complex retrieval method is likely to yield little information. 

The \textit{contrast effect} \citep{Cauley2018} between hotter and cooler regions of the star is exacerbated for TRAPPIST-1 by the presence of molecular absorption features in the star over the wavelengths probed by WFC3 G141, which in some ways mimic or counter the expected planetary signal. 
In this analysis we take the following steps to examine the measured transmission spectrum to attempt to mitigate the effects of stellar contamination on the real planetary spectrum of TRAPPIST-1g. 
\begin{enumerate}
\item Use the out-of-transit stellar flux measurements in each visit to create an average of the baseline star to determine a best fit stellar temperature considering a single stellar model. 
\item Following a similar method outlined in \citet{rackham2018} (and references therein), fit a multi-component stellar model to the out-of-transit stellar spectra to determine the contrast effect for different temperature regions and contamination fractions of the star occulted by the planet during transit. 
\item Determine the probability of the photospheric geometries with relation to the occulted stellar flux under the transit chord. 
\item Use the contrast effect computed from the contamination fractions and stellar flux for the different temperatures to correct the measured transmission spectrum in each wavelength channel. 
\item Use the transit depths, with corrections applied, to fit for a grid of planetary atmosphere models to place constraints on the planetary atmosphere. 
\end{enumerate}
We detail steps 1 and 2 in this section, with steps 3--5 detailed in \S\ref{sec:transmission_fit} and \S\ref{sec:interp} respectively. 

\subsection{Fitting the out-of-transit stellar spectra}\label{sec:star_fit}
To compare stellar models, in theoretical units, to observational data, in electron counts, we first transform the models into comparable units to the data. We use the Phoenix-COND stellar models \citep{Husser2013} to represent our star, TRAPPIST-1. We tested a series of other stellar models (e.g., BT-SETTL, BT-DUSTY, \citealt{Allard2012}; CIFIST, \citealt{Baraffe2015}; MARCS, \citealt{Gustafsson2008}). All models performed equally well at fitting the out-of-transit spectrum, but the Phoenix-COND models allow for self-consistent calculations of the stellar limb-darkening coefficients using the Phoenix-COND intensity profiles, as detailed in \S\ref{sec:ld}. 

We interpolate the stellar models onto a common 0.05\AA~ resolution; this is the approximate original wavelength resolution of the models, but regions of the models are at finer resolutions and placing it onto a uniform grid is necessitated by the need to map the spectra to a common wavelength solution of our data later in the process. We linearly interpolated from log(g)\,=\,5 and log(g)\,=\,5.5 to log(g)\,=\,5.22, binned up to 4$\times$ oversampled out-of-transit spectral resolution (i.e., 0.25\,pixel bins), convolved with the WFC3 center-of-fov (field of view) PSF at 1.4\,$\mu$m, and then binned to the resolution of the dataset. 
To match the units of the data -- electron counts -- the models (in units of erg/s/cm$^2$/\AA~ at the stellar surface) were converted to flux density at Earth (by a factor of (R$_{*}$\,/\,d)$^2$). We then converted to e$^-$/s/\AA~ through the 2011 WFC3 sensitivity curve\footnote{http://www.stsci.edu/hst/wfc3/documents/ISRs/WFC3-2011-05.pdf} (units e$^-$/s/\AA~ per erg/s/cm$^2$/\AA), and finally converted to pure electron counts by accounting for exposure time (112\,s) and the wavelength size of the spectral channels. 

It is important that the models be converted to the correct units for comparison to the data. This avoids arbitrary normalization which is not physically motivated, as rescaling the fluxes of the models can lead the belief that a model with an incompatible overall flux is an acceptable fit to the dataset. The use of the physical units allows for the potential breaking of any degeneracies between the stellar models.

For each visit we generate an average stellar spectrum from all of the out-of-transit exposures (excluding the first orbit, and the first exposure in each orbit as with the fit transit time series data). We use the measured average stellar spectrum out-of-transit to determine the best fit single stellar temperature and multi component models where,
\begin{equation}
F_{t} = F_0(1-X_{s1}-X_{s2}) + F_1X_{s1} + F_2X_{s2}\mathrm{.}
\label{eq:3tfluxfit}
\end{equation}
Here F$_t$ is the combined stellar counts, F$_0$, F$_1$, and F$_2$ are the electron counts of each different stellar model defined by their different temperatures, and X$_{s1}$ and X$_{s2}$ are the fraction of the star accounted for by each flux value. This can be set such that one, two and three component models are considered by fixing either one or both of $X_{s1}$ and $X_{s2}$ at 0.  

\begin{figure*}
\centering
\includegraphics[width=0.93\textwidth]{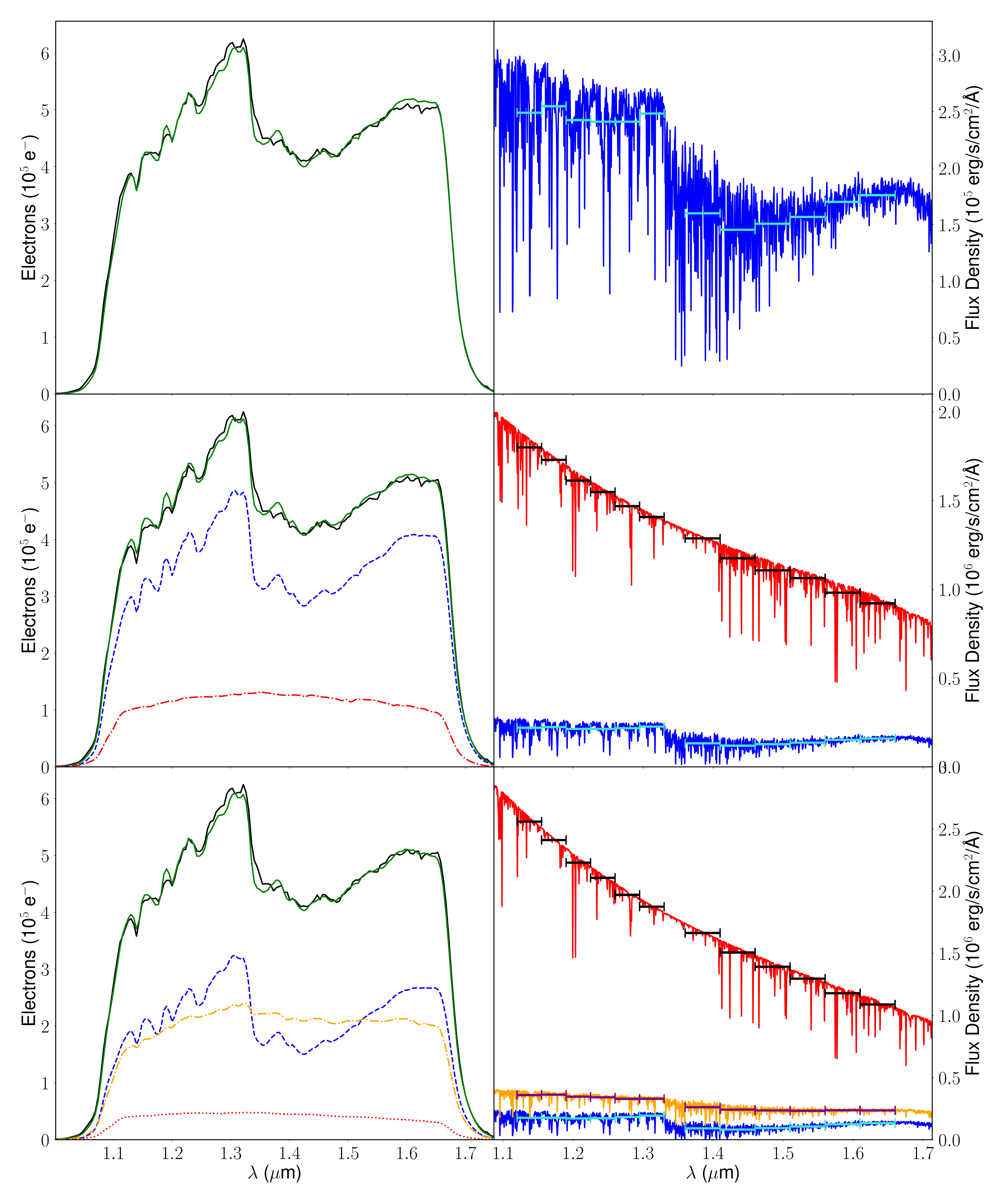}
\caption{Left: the average out-of-transit stellar spectrum measured using the WFC3 G141 grism (black) fit with one (top), two (middle), and three (bottom) stellar models with the resultant model in green. The uncertainty on the measured stellar spectrum is on the order of 700 electrons which is not visible on this scale. The main differences between each reconstruction and the data can be seen between 1.2--1.3\,$\mu$m and at $\sim$1.6\,$\mu$m. Right: stellar spectral models binned to 5\AA~ resolution, with the wavelength bins used for the transmission spectrum marked on each model. The low temperature spectra (blue) show a distinct molecular water feature at 1.4\,$\mu$m, which can also be seen in the measured spectrum (black). }\label{fig:stellar_spec}
\end{figure*}

\subsection{The out-of-transit model fits}
In general, for an inactive, quiet star, at low resolution a single stellar model can be used to represent the measured stellar photosphere. When considering an active star, a multi-temperature model will typically be adopted (e.g., \citealt{rackham2018,Morris2018,Ducrot2018}), to account for cold/hot regions on the star. For TRAPPIST-1, \citet{Morris2018} showed that the activity of the star as measured with \textit{Kepler} and \textit{Spitzer} was best described by a two temperature fit, where T$_{spot}\gtrsim$\,5300\,$\pm$\,200\,K with a fractional coverage of 0.4\%. However, this is a lower limit to the spot coverage as they note in their Appendix 1. Additionally, the study presented in \citet{Ducrot2018} also conclude that the photosphere of TRAPPIST-1 is most likely described by a base photosphere with small hot faculae ($>$4000\,K); however, they note that the star may also be fit with high latitude cold spots. Conversely, \citep{rackham2018} find larger spot covering fractions for TRAPPIST-1 of $f_{spot}$\,=\,8$^{+18}_{-7}$\% for much cooler temperatures of spot and faculae, which are fixed to scaled values of the photosphere (see \citealt{rackham2018} Table 1). In both of these cases the baseline stellar photosphere is fixed to T$_\mathrm{eff}\,\approx$\,2500\,K, with \citet{rackham2018} quoting a photospheric temperature of an M8V star, like TRAPPIST-1, as T$_\mathrm{eff}$\,=\,2400\,K. \citet{zhang2018} apply the temperature ranges quoted in \citet{rackham2018} to the measured stellar spectrum of TRAPPIST-1 when probing the effects of spot contamination on transmission spectra. However, we note here that the small fractional coverage ($\simeq$8\%) of an extremely cold ($\simeq$1900\,K) component lead to the case where the three temperature fit of \citet{rackham2018} and \citet{zhang2018} are, effectively, two-temperature fits, due to the $T^4$ scaling relation in stellar flux. 

In this analysis we fit for a one, two, and three temperature components to the stellar photosphere using the scaled Phoenix-COND models as described above. Following the description of \citet{rackham2018} these represent a pure photosphere, photosphere plus spot or faculae, and photosphere plus spot and faculae. We allow our two- and three-temperature fits to probe a much wider range of temperatures than those previously quoted. In all cases we fit the first temperature in the range 2300\,K $\leq$ T$_\mathrm{eff}\,\leq$\,3000\,K, typical of M4V--M9V stars, and fit for additional temperatures in the range 2300\,K\,$\leq$\,T$_\mathrm{eff}\,\leq$\,6000\,K. This allows for the possibility of colder spot temperatures (e.g., \cite{zhang2018} -- although we note their spot temperature posterior distribution is truncated, potentially suggesting a hotter spot temperature than quoted), as well as allowing for the hotter spot temperature found in \citet{Morris2018}. 

For each individual temperature we step through increments of 25\,K by interpolating between the Phoenix-COND models which are 100\,K apart. In each model we keep [Fe/H]\,=\,0.0, as higher metallicities were not available for some models and it is close to the accepted value of [Fe/H]\,=\,0.04 \citep{vangrootle2018}. Each stellar model in the two and three temperature fits has an associated stellar contamination fraction, i.e. how much of the stellar disk can be accounted for by each stellar model in each fit (see equation \ref{eq:3tfluxfit}). We fit these fractions in increments of 1\%, forcing the total fraction of temperatures to be in the range 0--100\%. 

However, we found that none of the fits -- one, two, or three temperature -- were an acceptable match to the data, producing reduced chi-squared statistics of the order of 3000 or more (Fig. \ref{fig:stellar_spec}). The best example of this is the single temperature fit. As our models are correctly converted to electron counts, an increase in effective temperature increases the flux, and thus electron counts, across all wavelengths. There is, therefore, a model with the correct overall electron count to fit the data. However, this model, T$_\mathrm{eff}\,\simeq$\,2450\,K, has an inverse water feature too large to match the measured spectrum. To correctly fit the \textit{relative} depth of the water feature, a higher temperature model would have to be employed -- and indeed, would produce an acceptable fit if the data were arbitrarily normalized, such as in the case of the fits employed by \citet{zhang2018}. The more rigorous treatment of the models applied in this analysis, where both models and data are in absolute units, leads to the case where this degeneracy is not broken, and thus we must consider other causes of this discrepancy.

One source of discrepancy between our data and model counts is the correction applied through the WFC3 sensitivity file. It is possible that there is an additional, unknown source of throughput correction that is not accounted for in conversion from flux density at Earth to e$^-$/s/\AA~. This additional loss between photons being received at Earth and electrons being recorded as detected would then mean our models overestimated the electron count.

The second potential cause of count offset is the conversion factor applied one stage before this -- the conversion from flux density at the stellar surface to flux density at Earth. This conversion requires a factor of (R$_{*}$\,/\,d)$^2$. $d$ is known to high precision, thanks to the excellent parallax as provided by the \textit{Gaia} Data Release 2 (DR2; \citealt{gaia2016,gaia2018}). However, R$_*$ is much less constrained. The scaling parameter therefore could be interpreted as an incorrect conversion from stellar surface to Earth. Thus our electron counts would need to be corrected by a factor of (R$_\mathrm{*,new}$\,/\,R$_*$)$^2$ to account for the assumption made that R$_*$\,=\,0.121 R$_\odot$ \citep{vangrootle2018}. This factor can be viewed as the correction from the literature stellar radius to the implicit model radius; it is well known that there are issues with theoretical models underestimating the radii of low mass stars (e.g., \citealt{Lopezmorales2007,Boyajian2012}).
\begin{table}
\centering
\caption{Updated TRAPPIST-1 stellar parameters from this study. }
\begin{tabular}{lcccc}
\hline
\hline
Paper & R$_*$/R$_\odot$ & M$_*$/M$_\odot$ & T$_{eff}$ & L$_*$/L$_\odot$ \\
\hline
V18 & 0.121 & 0.089 & 2516 & 0.000522  \\
D18 & 0.121 & 0.089  & 2511 & 0.000522  \\
F15 & 0.117 & 0.082 & 2557 & 0.000524  \\
\textbf{This work} & \textbf{0.117} & \textbf{0.080} & \textbf{2400} & \textbf{0.000523} \\ 
\hline
\multicolumn{5}{l}{V18 - \citet{vangrootle2018}; D18 - \citet{delrez2018}}\\
\multicolumn{5}{l}{F15 - \citet{Filippazzo2015}}\\
\end{tabular}
\label{table:star_params}
\end{table}

\begin{table*}
\centering
\caption{Scenarios considered for the occulted portion of the star by the transiting planet given three different reconstructions of the stellar flux.}
\begin{tabular}{lllllll}
\hline
\hline
Name & T$_{1}$ & T$_2$ & T$_3$ & Ratio & T$_{eff}$ & Comment\\
~ & (K) & (K) & (K) & (\%) & (K) &   \\
\hline
\multicolumn{7}{ c }{\textbf{One Stellar Temperature}}\\
1T & 2663 & - & - & 100 & 2663  & D$_M$ = D, A = 1 \\
\multicolumn{7}{ c }{\textbf{Two Stellar Temperatures}}\\
2T & 2563 & 5100 & - & 97:3 & 2652 & D$_M$ = D, A = 1 \\
2Tc & 2563 & - & - & 97 & 2563  & planet transits only lowest T$_\mathrm{eff}$ part of the star\\
2Th & 5100 & - & - & 3.5 & 5100 &planet transits only highest T$_\mathrm{eff}$ part of the star\\
\multicolumn{7}{ c }{\textbf{Three Stellar Temperatures}}\\
3T & 2400 & 3000 & 5825 & 64:35:1 & 2641 & D$_M$ = D, A = 1 \\
3Tc & 2400 & - & - & 64 & 2400 &  planet transits only lowest T$_\mathrm{eff}$ part of the star\\
3Tm & 3000 & - & - & 35 & 3000 & planet transits only middle T$_\mathrm{eff}$ part of the star\\
3Th & 5825 & - & - & 1 & 5825 & planet transits only highest T$_\mathrm{eff}$ part of the star \\
3Tc+m & 2400 & 3000 & - & 64:35 & 2609 & planet transits combination of lowest/middle T$_\mathrm{eff}$\\
3Tc+h & 2400 & 5825 & - & 64:1 & 2452 & planet transits combination of lowest/highest T$_\mathrm{eff}$\\
3Tm+h & 3000 & 5825 & - & 35:1 & 3080 & planet transits combination of middle/highest T$_\mathrm{eff}$\\
\hline
\end{tabular}
\label{table:correction}
\end{table*}
We therefore add an additional parameter to the fits to the out-of-transit spectrum: a scaling parameter, $\iota$. We choose to interpret the parameter as a physical radius scaling parameter, and thus actually consider the relative change in stellar radius, $\mathcal{R} = \sqrt{\iota}$. Our goodness-of-fit criterion is therefore $\chi^2 = \sum_i (dc_i - mc_i\times\mathcal{R}^2)^2\,/\,\sigma_i^2$ where $dc$ is the data electron count in the $i$th bin, $mc$ is the model count in the $i$th bin, and $\sigma$ is the uncertainty of the data counts, assumed to be described by Poissonian statistics. By interpreting this parameter as a scaling on the radius rather than an arbitrary scale factor, we use the data and models to explore a physically motivated phase space. 

\subsection{Information gained from the out-of-transit stellar flux}
We apply the one, two, and three temperature fits to the out-of-transit spectra to both visits in turn to determine if there is any time variable components that need to be considered. From this analysis we find that both observations, which were taken $\sim$1 year apart, are consistent with each other to within 3\% in spot temperature; 2\% in radius scaling parameter, $\mathcal{R}$, and hot spot coverage fraction; and 15\% in cold spot coverage fraction, giving on the order of 5\% spot variability, similar to the 3\% variability seen in the \textit{Spitzer} lightcurve \citep{delrez2018}.

In Fig.\ref{fig:stellar_spec} we show each of the stellar reconstructions and their components compared to the measured average stellar spectrum. Each of the spectra show the characteristic edges caused by the WFC3 G141 grism throughput, and the inverse water feature on the stellar photosphere. We also show the full resolution stellar models in their original units with the measured transmission spectral bins for comparison and further analysis in \S\ref{sec:transmission_fit}.

The single temperature fit requires a 2663\,K model with a relative radius factor of 0.84 (in the sense that the model radius is too small compared with the \citealt{vangrootle2018} value), with reduced chi-squared $\chi^2_\nu \simeq$\,340. Across both visits and all N-component stellar reconstructions, we consistently find a radius scaling factor of 0.84. We thus conclude that the implicit model radius must be of the order of 0.1\,R$_\odot$, consistent with that produced by the CLES suite of models \citep{vangrootle2018} for a $\sim$1\,Gyr 0.08\,M$_\odot$ star (see below).
The two-temperature fit agrees with \citet{Morris2018}, requiring a 2563\,K main photosphere with $X_{s1}$\,=\,3\% coverage of spots with temperature 5100\,K; the goodness-of-fit criterion is $\chi^2_\nu\,\simeq$\,265, a significant improvement over the one component model. Finally, our three-temperature fit finds a 2400\,K photosphere, 3000\,K spots covering $X_{s1}$\,=\,35\% of the star, while still requiring much hotter spots -- now 5825\,K -- covering $X_{S2}$\,=\,1\% of the star, with an improvement of the goodness-of-fit to $\chi^2_\nu \simeq$\,255. 

This criterion suggests a poor fit, which can clearly be seen in Fig.\ref{fig:stellar_spec}, especially at $\sim$\,1.15$\mu$m and $\sim$\,1.35\,$\mu$m. The poor fits are largely due to the models not necessarily reproducing fine details on a pixel-by-pixel basis, only instead reproducing broader features of the measured spectrum. This, in addition to necessary interpolation of the stellar models, leads us to limit our analysis of the out-of-transit spectrum, simply quoting the best fit model. The model grid of stellar spectra step in values of 100\,K, we therefore adopt this uncertainty on the stellar temperatures quoted, however, do not derive uncertainties on the fractional coverage or correction factors based on this value. 

This poor fit between model and data is also seen in \citet{zhang2018} where they state that the uncertainties of their spectra must be inflated by a factor of 23 to fall into agreement with their model fits. This significant inflation factor applied to the data is more likely interpreted as informing on the rejection of the \textit{model}. Requiring an uncertainty inflation of 23 to get reduced chi-squared statistics of the order 1 ($\chi^2 \sim$\,140 on 135 degrees of freedom, table 11, \citealt{zhang2018}) suggests an ``uncorrected'' reduced chi-squared of approximately 530, over twice as poor a fit as our best fitting three-temperature fit.

We consider the effect that each stellar fit has on the contrast effect and thus the transmission spectrum of TRAPPIST-1g in Section \ref{sec:transmission_fit}, but we briefly discuss some of the implications the out-of-transit spectrum fit has on TRAPPIST-1 itself here. We first consider the single temperature fit, the most frequently assumed, and the simplest, model. We require an effective temperature of 2663\,K to fit the out-of-transit spectrum and match the inverse water feature observed. This temperature is $\sim$150\,K hotter than the literature value for the star (2511\,K; \citealt{vangrootle2018}), and over 250\,K hotter than the typically quoted photospheric effective temperature for an M8V star (2400\,K; \citealt{Kaltenegger2009,Filippazzo2015}). Forcing the fit to T$_\mathrm{eff}$ = 2511\,K still requires a minor radius deflation ($\sim$5\%), but results in a fit over five times worse, with the colder temperature unable to fit the relative size of the water feature. The hotter photosphere would suggest that TRAPPIST-1 is an M5V-M6V star, requiring a mass of the order 0.11 M$_\odot$ \citep{Kaltenegger2009}, far higher than any previous estimates of the mass of the star. We therefore rule out the single-temperature fit based on the discrepancies in the out-of-transit spectrum fits.

Given the nature of M stars and the measured variability seen with \textit{Spitzer} and \textit{Kepler} it is likely that there are active regions on the star, and thus likely that a multi-component fit is physically justified. In \S\ref{sec:transmission_fit} we discuss how the planet can be used to further rule out/in each case. 
The main difference between the out-of-transit spectrum we present here and the analysis done in the recent literature is the photospheric effective temperature. The star is quoted as having a single temperature $\simeq$\,2500\,K (e.g., \citealt{rackham2018}). However, we find a main photospheric temperature of 2400\,K, in agreement with the relationship between spectral type and $T_\mathrm{eff}$. This effective temperature suggests an old (consistent with \citealt{Burgasser2017}) 0.08 M$_\odot$ star.

Combining the mass with the direct measurement of the density (51.1 $\rho_\odot$, \citealt{delrez2018}) gives 0.117 R$_\odot$, giving stellar parameters consistent with \citet{Filippazzo2015}. This radius implies that the theoretical models underestimate the stellar radius by 15\%, consistent with previous studies. Additionally, a simple coverage-weighted luminosity calculation (using the self-consistent model radius) for the three-temperature fit gives $\log(L / L_\odot) = -3.27$, consistent with the luminosity quoted by \citet{vangrootle2018} to within 1$\sigma$. We therefore conclude this section by summarizing TRAPPIST-1 as a 2400\,$\pm$\,100\,K, 0.08\,M$_\odot$, 0.117\,R$_\odot$, M8V star (see Table \ref{table:star_params}), with $\sim 35\%$ 3000\,$\pm$\,100\,K spot coverage and a very small fraction, $<3\%$, of $\sim$ 5800\,$\pm$\,100\,K hot spots, which has negligible changes to the stellar limb-darkening coefficients. This change in stellar radius will have an impact on the measured planetary radius via the transit depth, and therefore the bulk density of the planet for a given mass. We discuss these implications in \S\ref{sec:transmission_fit}. 

\section{The effect of the star on the planetary transmission spectrum}\label{sec:transmission_fit}
We can use the transiting planet to evaluate the plausible scenarios for the stellar photosphere by using the measured transmission spectrum to approximate the region of the star being occulted by the planet.  
During a planetary transit the occulted portion of the star is assumed to be consistent with the unocculted star measured out-of-transit. However, if the star has inhomogeneities on the surface such as spots or faculae that are not present in the shadow of the planet, the occulted portion of the star will be on average different to the unocculted star. Thus, imprints from the stellar atmosphere will be seen in the transmission spectrum, especially if the star has a strong wavelength dependence as TRAPPIST-1 does.  

\begin{figure}
  \centering
  \includegraphics[width=0.45\textwidth]{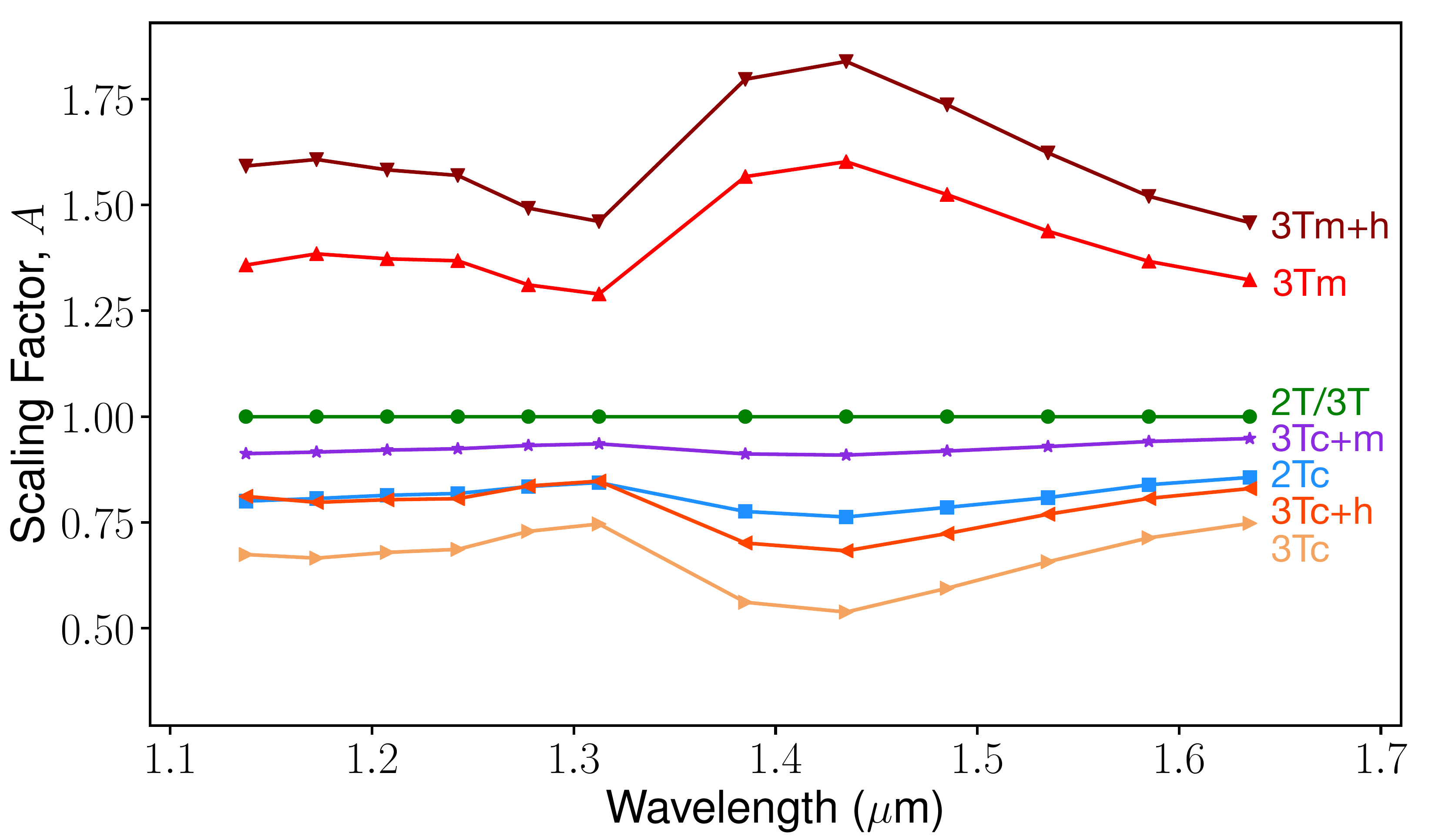}
  \caption{The wavelength dependent correction factor for seven different scenarios associated with the portion of the star being transited by the planet (see Table \ref{table:correction}).}\label{fig:A_factor}
\end{figure}

As detailed in \S\ref{sec:star_fit}, the stellar contamination caused by portions of the star being darker or brighter than the base star can be equated to a fraction of the star exhibiting a different stellar flux compared to that of an assumed ``clean'' stellar spectrum. 
As stated in \citet{rackham2018} the contamination factor, here denoted as $A$, will be a multiplicative factor on the measured transit depth such that;
\begin{equation}
\left(\frac{R_p}{R_*}\right)_{M}^2 = A \times \left(\frac{R_p}{R_*}\right)^2 \mathrm{,}\\
\end{equation}
\begin{equation}
\mathrm{or ~~} D_M = A \times D \mathrm{,}
\end{equation}
where the $M$ denotes the measured value, $D$ is the planet-to-star radius ratio, and
\begin{equation}
A = \frac{F_{x}}{F_{0}(1 - X_{s1} - X_{s2}) + F_{1}X_{s1} + F_{2}X_{s2}}\mathrm{,}
\label{eqn:A}
\end{equation}
which is a scaling factor on the real planet-to-star radius ratio normally assumed to be 1. Here F$_x$ represents the flux of the chosen stellar component being occulted by the planet during transit and can be a linear combination of any of the individual stellar fluxes considered and their associated fractional coverage. As $A$ is a unit-less scaling factor it is important to use the flux for each of the components in the same unnormalized units of flux density (see Fig.\ref{fig:stellar_spec}) for each wavelength dependent bin associated with the measured transmission spectrum. 

\begin{figure}
\centering
\includegraphics[width=0.45\textwidth]{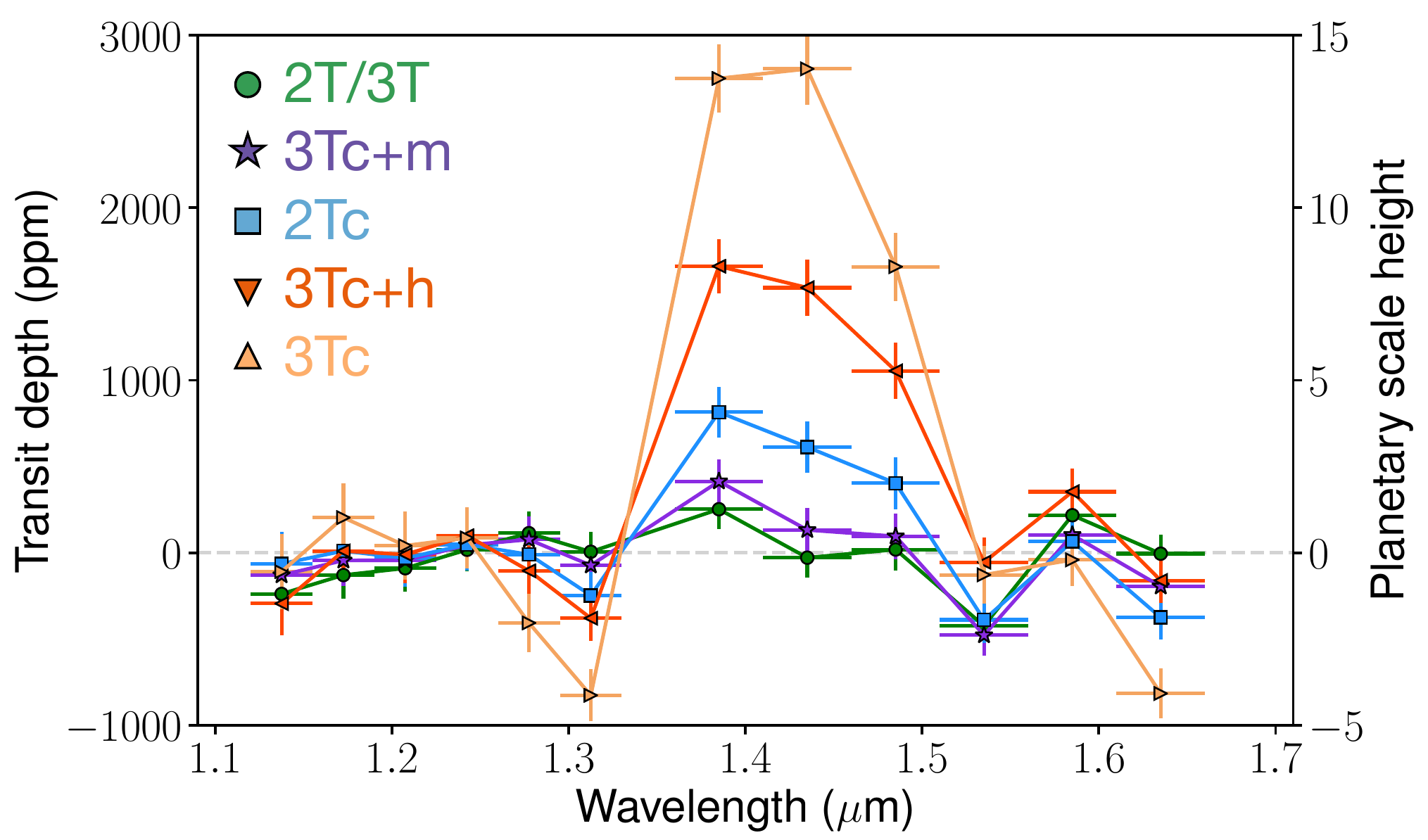}
\caption{The corrected transmission spectrum of TRAPPIST-1g for five different scenarios of the \textit{contrast effect} associated with the occulted starlight. From these we can rule out scenarios 3Tc, 3Tc+h, and 2Tc as they extend to unrealistic scale heights (i.e. $>$\,5H) for the expected planetary atmosphere. All spectra have been individually median subtracted such that they all align around zero for direct comparison.}\label{fig:all_real_transmission}
\end{figure}
\subsection{Application to the measured transmission spectrum}
Previous studies have calculated the effective stellar contamination effect (e.g., \citealt{rackham2018,zhang2018,Morris2018b,Ducrot2018}) to make comparisons to the stellar models. In this study we take this one step further and use the stellar models and the computed correction factor in each wavelength bin to correct the measured transmission spectrum and obtain the spectrum from the planet alone. 

The three stellar reconstructions described in \S\ref{sec:star_fit} result in 11 geometric scenarios for the occulted portion of the star during the transit of TRAPPIST-1g. We describe each of these scenarios in this section and their plausibility and implications for the star and planet in turn. For each scenario the wavelength dependent correction factor will depend upon the portion of the stellar flux being transited. In cases where the planet is assumed to transit two or three different stellar components we assume they are in the same ratio as the full unocculted star. For example, in the three-temperature fit where the ratio is 35:64:1 if we assume the planet transits only the first two components we apply them at a ratio of 35:64 to calculate $A$. Each of the scenarios are listed in Table \ref{table:correction}.

Three of the scenarios (1T/2T/3T) result in the average flux of the occulted portion of the star equal to the average unocculted stellar flux. In these scenarios $A\,=\,1$ at all wavelengths as no stellar contamination is present in the planetary spectra. In each of these, the measured transit depth $D_M\,=\,D$, and will not contain any \textit{contrast effect} from the star. All of these scenarios result in average stellar temperatures with around 2650\,K and a $\Delta$T$_\mathrm{eff}$\,=\,22\,K; we therefore use a single temperature of T$_\mathrm{eff}$\,=\,2663\,K to compute the limb-darkening to obtain the measured planetary transmission spectrum (see Table \ref{table:transmission_spec}). 
However, given our constraints based on the star we can rule out the 1T case being the favored description of the system (see \S\ref{sec:stellar}). This leaves 2T and 3T as plausible scenarios for the star planet combinations, which in turn require the transit chord to occult the star in the same contrast ratio as the out-of-transit star, i.e. the stellar flux ratios are 97:3 and 64:35:1 under the planet shadow for 2T and 3T respectively for the given temperatures listed in Table \ref{table:correction}. Based on these scenarios we can approximate the size of the smallest flux contribution within the occulted portion of the star and thus the plausibility of these scenarios given no occulted spot features were seen in the transit lightcurves. In the 2T and 3T cases (R$_{spot}$/R$_*$)$^2$\,=\,0.03 and 0.01 respectively. This translates to a spot size within the transit chord with a physical radii of $\leq$\,1.22 megameters (Mm); in contrast the smallest spots on the Sun are $\sim$1.75\,Mm \citep{Solanki2003}. This is still potentially plausible for TRAPPIST-1 given it is an M star and small-scale magnetic activity could potentially be present to this level. We also note that this is below the precision obtained with our transit time series measurements and thus still plausible even though we do not see any evidence of spot crossings. These scenarios follow the similar conclusions drawn in \citep{Morris2018b}, which show that the measurements based on the \textit{Spitzer} data suggest that the planets all transit the mean photospheric value without being able to rule out small-spot crossings where (R$_{spot}$/R$_*$)$^2$\,$<$\,0.04.  
\begin{figure}
\centering
\includegraphics[width=0.45\textwidth]{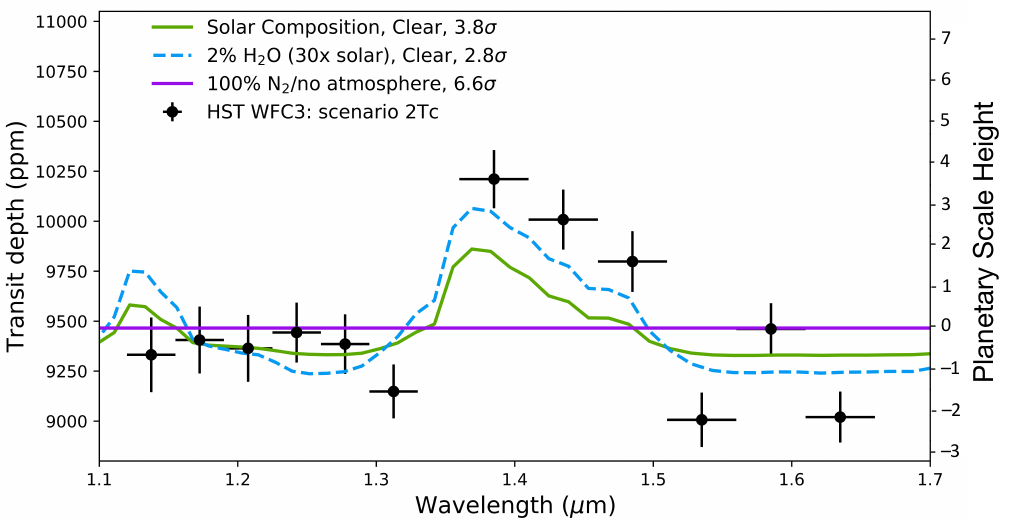}
\caption{Transmission spectrum for scenario 2Tc, demonstrating that the scenarios 3Tc and 3Tc+h can also be ruled out due to unrealistic planetary atmospheric scenarios required to fit the large absorption features.}
\label{fig:transmission_2Tc}
\end{figure}

In two scenarios (2Th and 3Th) the planet is required to transit only a region of the star that has T$_\mathrm{eff}\,>\,$5000\,K which covers $\,<\,$4\% of the stellar disk. This is geometrically implausible, as it would require the hot portion of the star to be aligned exactly along the transit chord with no breaks, therefore we do not consider these two scenarios in our analysis. This leaves six remaining scenarios which need to be considered: one where the planet transits just the colder part of a two component fit (2Tc), two where the planet transits a single component of a three component fit (3Tc and 3Tm), and three where the planet transits a combination of two components in a three component fit (3Tc+m, 3Tc+h, and 3Tm+h). Each scenario is summarized in Table \ref{table:correction}. 

For each of the seven scenarios to consider we calculate the wavelength dependent scaling factor, $A$, shown in Fig.\,\ref{fig:A_factor}. Based on these values alone it is clear that scenarios 3Tm and 3Tm+h, which assume the planet is transiting the middle temperature region alone or a mixture of the middle and hotter temperature combined respectively, can be ruled out as they will further enhance the contribution of the stellar absorption feature onto the transmission spectrum. We discuss the implications of the five remaining scenarios in the following section after correcting the measured transmission spectrum by the respective $A$ values in each wavelength bin for each scenario to obtain the uncontaminated planetary spectrum.

\begin{figure*}
\centering
\includegraphics[width=0.95\textwidth]{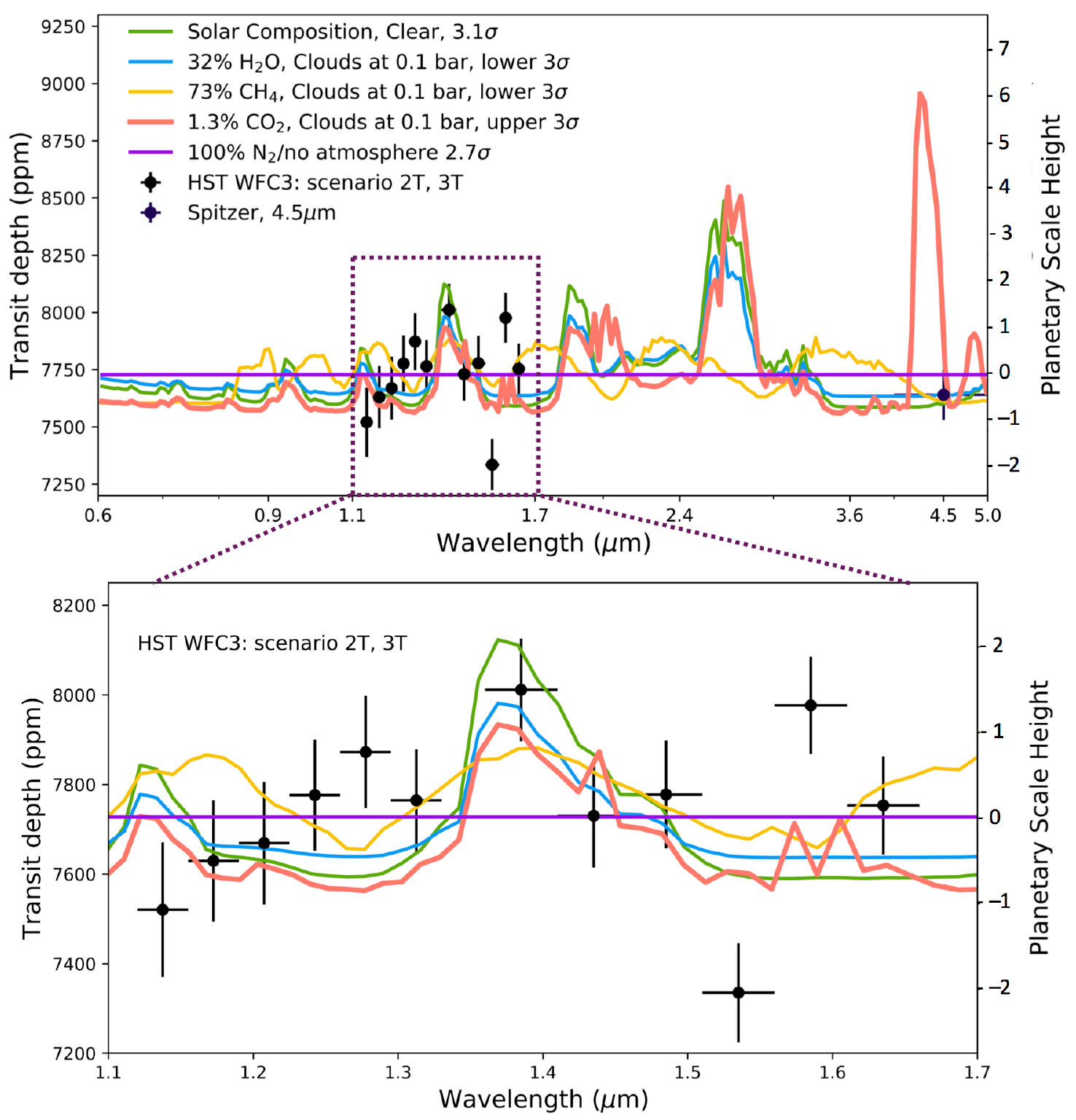}
\caption{The most likely transmission spectrum scenario for TRAPPIST-1g assumes no stellar contamination is present in the measured transit depths (scenarios 2T and 3T). We also include the \textit{Spitzer} 4.5\,$\mu$m measurement presented in \citet{delrez2018} under the same assumption. Under this assumption, over the wavelengths probed, we are able to rule out a clear solar H$_2$/He-dominated atmosphere at 3.1-$\sigma$ and place an lower bound on the H$_2$O content to 32\%. The sigma confidence values listed state by how much each model can be ruled out. Spectroscopic observations at IR wavelengths, especially beyond 2.4\,$\mu$m will be essential to distinguish between the different models and molecular content for the atmosphere of TRAPPIST-1g as any current additional structure is likely statistical scatter at the resolution and precision measured. Here one planetary scale height H\,=\,200\,ppm (see text for details).}
\label{fig:transmission_2T3T_inset}
\end{figure*}
\section{Interpreting the planetary transmission spectrum} \label{sec:interp}
Out of 11 possible combinations of stellar reconstructions and planetary geometries only six remain physically plausible (2T and 3T, 3Tc+m, 2Tc, 3Tc+h, 3Tc). For each of these scenarios we use the computed $A$ values to correct the measured transmission spectrum, where for scenarios 2T and 3T, A\,=\,1 at all wavelengths. The resultant planetary transmission spectrum is shown in Fig.\,\ref{fig:all_real_transmission} for each of these scenarios described in Table \ref{table:correction} and \S\ref{sec:transmission_fit}.  For this study, we approximate one planetary scale height for TRAPPIST-1g\,$\approx$\,200\,ppm\,$\approx$\,100\,km. Note that a robust determination of the planetary scale height will require future observations over an expanded wavelength range that captures more atmospheric features at higher resolution. Based on atmospheric models of TRAPPIST-1g the maximum extent of the planetary atmosphere in the WFC3 G141 wavelength range is $\approx$5\,H (e.g., \citealt{seager2010book,Burrows2014}). From this we can further rule out scenarios 3Tc, 3Tc+h, and 2Tc as they require planetary signals with unrealistic scale heights given T$_{eq}$\,=\,195\,K and g$_p$\,=\,7.4032\,ms$^{-2}$ (see Table \ref{table:system_params}). We show justification for this assessment in Fig.\,\ref{fig:transmission_2Tc} where the transmission spectrum for scenario 2Tc is plotted with a series of clear solar composition models (detailed in \S\ref{sec:planetary_models}). Fig.\,\ref{fig:transmission_2Tc} demonstrates the maximum amplitude of the planetary signal given the planetary scale height and an assumption of a H/He-dominated atmosphere for TRAPPIST-1g. As we are able to rule out scenario 2Tc, we are also able to confidently rule out scenarios 3Tc and 3Tc+h, which also require unrealistic amplitudes for the planetary signal from TRAPPIST-1g.

This leaves two remaining planetary transmission spectra to investigate (2T and 3T, 3Tc+m). One set of scenarios exist in which the measured transmission spectrum does not contain any additional contamination from the star (2T and 3T); the other scenario involves the planet transiting a homogeneous mix of two stellar flux components at T$_\mathrm{eff}$\,=\,2400 and 3000\,K at a ratio of 64:35 respectively, where the star has a third unocculted stellar flux component with T$_\mathrm{eff}$\,=\,5875\,K over 1\% of the star (3Tc+m). We note that the second scenario (3Tc+m) results in a correction factor $A\,\approx\,$0.9--0.97, which is similar to the value presented in \citep{Morris2018b} of A $\sim$ 0.84 (in a range of 0.82--1.04). 

\subsection{TRAPPIST-1g planetary models}\label{sec:planetary_models}
For each of the remaining scenarios (2T and 3T, 3Tc+m) we fit the planetary transmission spectrum with atmospheric models specific to TRAPPIST-1g. 
Each model, outlined in \citet{Batalha2018} and  \citet{Moran2018}, is based on a modified version of \texttt{CHIMERA} \citep{Line2013}, which is a one-dimensional correlated-\textit{k} radiative transfer code. It employs a 5-parameter double-gray analytic 1-D temperature pressure profile \citep{Guillot2010}, which, for non-irradiated systems, is approximately $T_z^4\,\sim\,0.75\times T^4 (p + 2/3)$, where $p$ is the height-dependent temperature pressure and $T$ is the equilibrium temperature. We implement a gray opacity source at specified pressures to approximate the effects of high opacity clouds, and the scattering parameterization of \citet{Lecavelier2008} to introduce a scattering cross section. We include in our models molecular opacity due to H$_2$/He collision-induced absorption, methane, water, carbon dioxide, and molecular nitrogen \citep{Freedman2008,Freedman2014}. We assume chemical abundances are constant with altitude (similar to \citealt{Batalha2018}). These simplified assumption about TRAPPIST-1g's atmosphere are motivated by the narrow wavelength range and relatively low SNR spectra explored here. To generate our atmospheric models we use a planetary mass of 1.148\,M$_\oplus$ and base planetary radius of 1.154\,R$_\oplus$.   

For each plausible spectrum, we explore H$_2$/He atmospheres with varying levels of either H$_2$O, CH$_4$, or CO$_2$, in addition to the null-hypothesis of a featureless spectrum or airless body. Although the data quality is not sufficient to do a full Bayesian retrieval, we find the limiting cases for atmospheric scenarios that can be ruled out to 3-sigma confidence. We do this by increasing the percentage of either H$_2$O, CH$_4$ or CO$_2$, from solar values, until a 3-sigma level is obtained. We also explore the effect of decreasing the pressure level of a grey cloud. 

\begin{figure}
\centering
\includegraphics[width=0.45\textwidth]{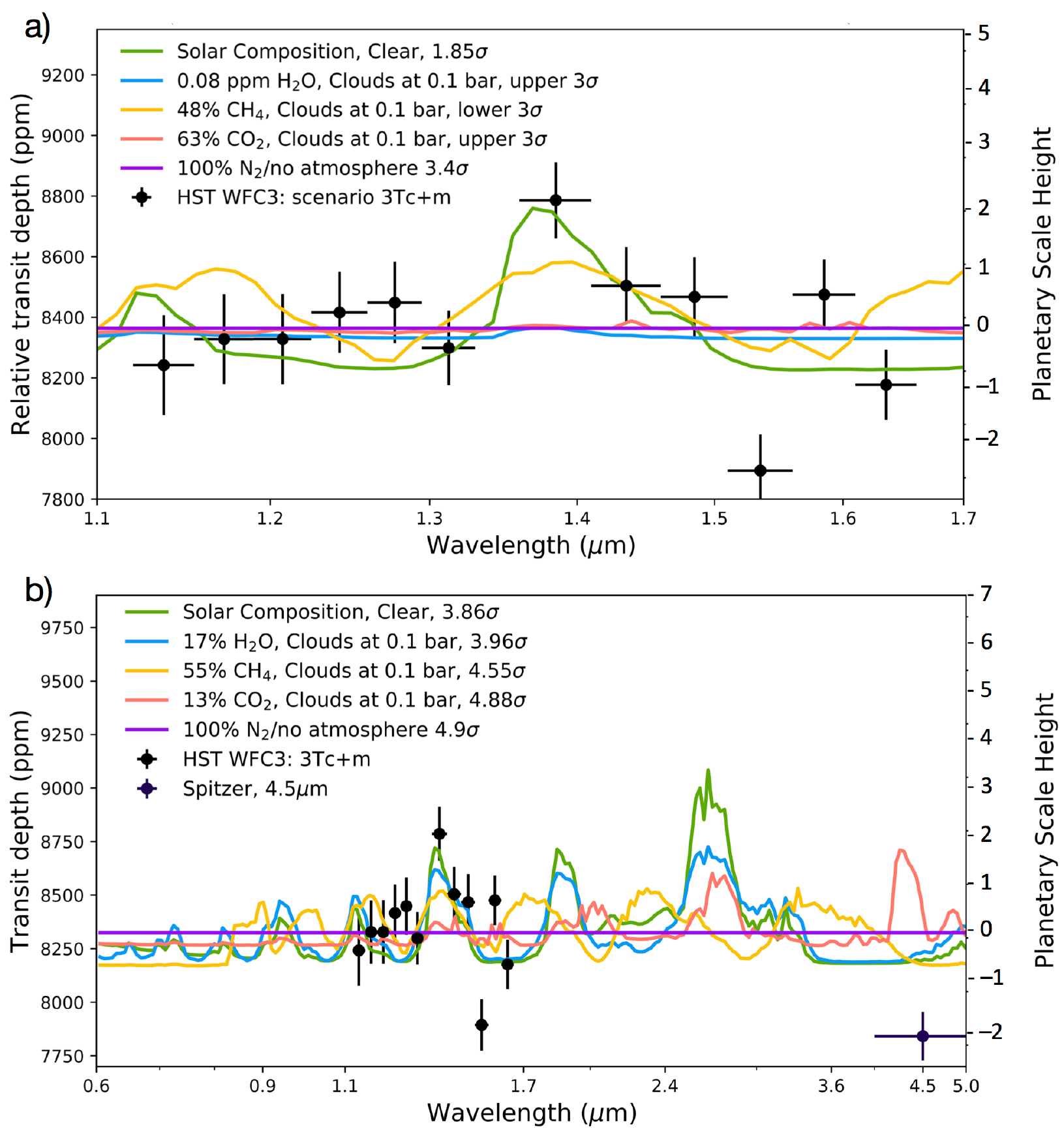}
\caption{Model analysis for scenario 3Tc+m, where the measured transit depth is corrected for stellar contrast effects to obtain the planetary transmission spectrum (see \S\ref{sec:transmission_fit}). a) Using the WFC3 data only we are able to rule out the null-hypothesis of a featureless transmission spectrum, either due to spectroscopically inactive gases over the observed wavelengths or an airless body, at 3.4-$\sigma$. b) We correct the \textit{Spitzer} 4.5\,$\mu$m value for the same stellar contrast effect and conduct the model analysis. Given the addition of the \textit{Spitzer} value we are able to rule out all considered model scenarios at greater than 3-$\sigma$ confidence. The WFC3 wavelength errorbars are hidden in the data points. 
The sigma confidence values listed in each panel state by how much each model can be ruled out.}
\label{fig:3Tcm_transmission}
\end{figure}

We present the transmission spectrum, assuming zero contrast effect from the star (scenarios 2T and 3T), in Fig.\,\ref{fig:transmission_2T3T_inset}. Under this assumption, we can exclude a solar composition clear atmosphere to 3.1$\sigma$, but we are unable rule out the null hypothesis of no atmosphere to this certainty, reaching a confidence level of only 2.7$\sigma$. 
We find that mixing ratios of 32\% H$_2$O or 73\% CH$_4$ produce 3$\sigma$ exclusions to the transmission spectrum as lower limits.
For carbon dioxide, we find an upper limit of 1.3\% CO$_2$, after which the data is not sufficient to rule out smaller scale height models. Due to the broad wavelength coverage of the photometric \textit{Spitzer} 4.5\,$\mu$m measurement \citep{delrez2018} we are unable to place strong constraints on the carbon content of the atmosphere. 

Correcting the measured transmission spectrum for the contrast effect associated with scenario 3Tc+m results in slightly larger spectral features at 1.4\,$\mu$m and an overall depth increase of $\sim$500\,ppm. Figure\,\ref{fig:3Tcm_transmission} shows the resultant transmission spectrum and model analysis. Conducting model analysis on the WFC3 data alone allows us to rule out the null-hypothesis of a featureless spectrum, or airless body, at 3.4-$\sigma$. However, when we additionally correct the \citep{delrez2018} Spitzer measurement based on a contrast effect for the same stellar scenario, where $A\,=\,0.9725$ between 4.0--5.0\,$\mu$m, we find that all atmospheric models can be ruled out at greater than 3-$\sigma$, suggesting that this scenario can also be discounted.

For either scenarios 2T and 3T, based on WFC3 measurements alone we cannot distinguish between the various allowed high mean molecular models within the precision of the current data.
While the wavelength region covered by \textit{HST} WFC3 G141 (1.1--1.7\,$\mu$m) cannot be used in this instance to definitively distinguish between planetary atmospheric compositions, we note that it is a useful wavelength range to use the planets to probe the stellar photosphere using the stellar absorption feature and geometry of the transit. 
Further precision, as will be possible with a reasonable amount of \textit{James Webb Space Telescope} time \citep{Morley2017,Batalha2018}, will be necessary to distinguish between carbon dioxide, water, methane, or nitrogen dominated atmospheres with or without clouds. In Figs. \ref{fig:transmission_2T3T_inset} and \ref{fig:3Tcm_transmission}b we show the full planetary transmission spectra from 1 to 5\,$\mu$m covered by the JWST NIRSpec prism. Any strong molecular features due to carbon dioxide, water, and methane, if present in the planetary atmospheres, would be seen in this wavelength region, especially beyond 2.4\,$\mu$m, and provide much better diagnostics regarding the composition of the atmosphere.

\section{Conclusion} \label{sec:end}
We present an analysis method to disentangle the planetary transmission spectrum from stellar molecular features using the out-of-transit stellar spectra, planetary transit geometries, and planetary atmospheric models. This method is especially applicable to late type M dwarfs over the WFC3 G141 grism wavelength range where these types of stars have significant molecular absorption features in their atmosphere. 
We use TRAPPIST-1g, the largest planet in the TRAPPIST-1 system, as a test case for this method, based on HST WFC3 G141 transmission spectra. We present a self-consistent analysis for the transmission spectrum of TRAPPIST-1g using two transit observations from the HST WFC3 G141 grism between 1.1--1.7\,$\mu$m. We use the out-of-transit stellar spectrum to fit the star with stellar models from the Phoenix-COND grid to determine the fraction of the star effected by potential active regions. We then apply the stellar contrast effect to correct for contamination on the transit depths to determine the true planetary transmission spectrum. 

From the analysis of TRAPPIST-1g we find the following:
\begin{itemize}
\item The out-of-transit stellar spectrum for TRAPPIST-1 can be best fit with stellar models corresponding to three temperature components at T$_{eff}$\,=\,2400, 3000, 5825\,K for a coverage fraction of 64:35:1 respectively (scenario 3T).
\item We find that TRAPPIST-1 is a 0.08 M$_*$, 0.117 R$_*$, M8V star with a photospheric effective temperature of 2400\,$\pm$\,100\,K, in agreement with \citet{Filippazzo2015}.
\item Both observations, taken $\sim$1 year apart, result in the same stellar model fits with a corresponding activity level of 5\%, similar to the measured 3\% variable activity measured in the \textit{Spitzer} and K2 data.
\item Given the determined stellar radius, we calculate a planetary radius of 1.124\,R$_\oplus$ from our joint broadband fit, which is on the lower edge of the 1-sigma bound presented in \citet{delrez2018}. Taking the mass of TRAPPIST-1g from \citet{Grimm2018}, we recalculate the planetary density of TRAPPIST-1g to be $\rho_p$\,=\,0.8214\,$\rho_\oplus$.  
\item Using the combination of stellar models and the geometry of the planetary transit we are able to rule out 8 of 11 geometric scenarios. These scenarios consider potential combinations of one, two, and three temperature components of the stellar photosphere that may be occulted during the transit. (see \S\ref{sec:transmission_fit}).
\item Out of the three remaining scenarios for the planet and star, two result in no contrast effect being measured such that the measured transmission spectrum is of the planet alone with no contamination by stellar spectral features. Based on the analysis of this planetary transmission spectrum we can rule out the presence of a solar cloud free H/He-dominated atmosphere at 3-sigma.
\item We are able to rule out the final plausible scenario for the planet and the star (3Tc+M) by including the \textit{Spitzer} 4.5\,$\mu$m measurements from \citep{delrez2018}. This scenario requires the transmission spectrum to be corrected for a minor contrast effect due to unocculted bright flux on the star. However, inclusion of the measured \textit{Spitzer} value allows us to rule out all model analysis at greater than 3-$\sigma$.
\end{itemize}

In summary, for the case of TRAPPIST-1g, we find that the planetary transmission spectrum is not likely contaminated by any stellar spectral features, with a clear solar composition H$_2$/He-dominated atmosphere ruled out at greater than 3-$\sigma$. The most likely scenario for the stellar photosphere is that of a three component flux model with a small fraction of flux, 1\%, potentially caused by magnetic activity. 

The WFC3 G141 wavelength range from 1.1--1.7\,$\mu$m is a useful probe of the stellar photosphere using the occulting planet to disentangle the most plausible stellar component geometries; however, it is not the most distinguishing wavelength for the planetary atmosphere. Combining this information with longer wavelength spectroscopic observations will be important to fully disentangle the effect of the star on the measured planetary spectrum. 



\section{acknowledgments}
We thank the anonymous reviewer for their useful comments on the manuscript. This work is based on observations made with the NASA/ESA Hubble Space Telescope that were obtained at the Space Telescope Science Institute, which is operated by the Association of Universities for Research in Astronomy, Inc. These observations are associated with programs GO-14873 and GO-15304 (PI. J. deWit). We thank and GO-15304 collaboration for their comments on this work, in particular Jeremy Leconte, Amaury Triaud, Sean N. Raymond and Valerie Van Grootel for their encouraging words.
This research has made use of NASA's Astrophysics Data System, and components of the IDL astronomy library, and the Python modules SciPy \citep{scipy}, NumPy, Matplotlib \citep{matplotlib}, and corner.py \citep{corner}. 

H.R.Wakeford acknowledges support from the Giacconi Fellowship at the Space Telescope Science Institute, which is operated by the Association of Universities for Research in Astronomy, Inc. J.Fowler received support through GO-15304 and G.Bruno received support through GO-14767; provided by NASA through a grant from the Space Telescope Science Institute, which is operated by the Association of Universities for Research in Astronomy, Incorporated, under NASA contract NAS.
S.E.Moran was supported in part by a Johns Hopkins University Catalyst Award. V. Bourrier acknowledges the National Centre for Competence in Research PlanetS supported by the Swiss National Science Foundation (SNSF), and has received funding from the European Research Council (ERC) under the European Union's Horizon 2020 research and innovation programme (project Four Aces; grant agreement No 724427).

Author contributions: The team was led by H.R. Wakeford. Data analysis was conducted by H.R.Wakeford, J.Fowler, G.Bruno. T.J.Wilson led the analysis of the stellar spectra along with J.Valenti and G.Bruno. J.Valenti and J.Filippazzo worked on the limb-darkening. S.E.Moran produced planetary model spectra and led the planetary interpretation with the help of H.R.Wakeford, N.E.Batalha and N.K.Lewis. N.K.Lewis and J.deWit planned the observations. H.R.Wakeford led the writing of the manuscript, and all authors provided comments on the submitted work.

\bibliography{ref_T1g}

\end{document}